\documentclass{aa}
\usepackage[varg]{txfonts}
\usepackage{natbib}
\bibpunct{(}{)}{;}{a}{}{,} 

\usepackage{xcolor}
\usepackage{amssymb}
\usepackage{amsmath}
\usepackage{CJK}
\usepackage{ulem}
\usepackage{booktabs}
\usepackage{epstopdf}

\newcommand{\ssst}{\scriptscriptstyle}
\newcommand{\E}[1]{\times 10^{#1}}

\newcommand{\chisq}{\chi_\nu^2}

\newcommand{\RAdot}[4]{{#1}^{{\rm h}}{#2}^{{\rm m}}{#3}\fs{#4}}
\newcommand{\decldot}[4]{{#1}^{\circ}{#2}'{#3}\farcs{#4}}

\newcommand{\s}{\,{\rm s}}      \newcommand{\ps}{\,{\rm s}^{-1}}
    \newcommand{\Msun}{M_{\odot}}
\newcommand{\cm}{\,{\rm cm}}    \newcommand{\km}{\,{\rm km}}

\newcommand{\erg}{\,{\rm erg}}        \newcommand{\K}{\,{\rm K}}
    
     \newcommand{\g}{\,{\rm g}}

        \newcommand{\NH}{N_{\ssst\rm H}}

\newcommand{\Rs}{R_{\rm s}}
\newcommand{\kTc}{kT_{\rm c}}
\newcommand{\kTh}{kT_{\rm h}}
\newcommand{\kTi}{kT_{\rm i}}
         
\newcommand{\nH}{n_{\ssst\rm H}}

\newcommand{\Chandra}{{\sl Chandra}}
\newcommand{\du}{d_{10}}

\newcommand{\snr}{W49B}

\begin{document}

\title{Asymmetric Type-Ia supernova origin of W49B as revealed from spatially resolved X-ray spectroscopic study}

\author{
Ping Zhou \inst{1,2}
\and
Jacco Vink\inst{1,3} 
}

\institute{Anton Pannekoek Institute, University of Amsterdam, PO Box 94249, 1090 GE Amsterdam, The Netherlands; \email{p.zhou@uva.nl; j.vink@uva.nl}
\and School of Astronomy and Space Science, Nanjing University,
Nanjing~210023, China
\and GRAPPA, University of Amsterdam, PO Box 94249, 1090 GE Amsterdam, The Netherlands}

\abstract{The origin of the asymmetric supernova remnant (SNR) W49B has been a matter of debate: 
is it produced by a rare jet-driven core-collapse supernova, or by a normal supernova that is strongly shaped by its dense environment?
Aiming to uncover the explosion mechanism and origin of the asymmetric, centrally filled
X-ray morphology of W49B, we have performed  spatially resolved X-ray spectroscopy and a search for potential point sources. We report new candidate point sources inside W49B.
The Chandra X-ray spectra from W49B are well-characterized by two-temperature gas components 
($\sim 0.27$~keV + 0.6--2.2 keV). The hot component gas shows a large temperature gradient 
from the northeast to the southwest and is over-ionized in most regions with 
recombination timescales of  1--$10\times 10^{11}$~cm$^{-3}$~s.
The Fe element shows strong lateral distribution in the SNR east,
while the distribution of Si, S, Ar, Ca is relatively smooth and 
nearly axially symmetric.
Asymmetric Type-Ia explosion of a Chandrasekhar-mass white dwarf 
well-explains the abundance ratios and 
metal distribution of W49B, whereas a jet-driven explosion and
normal core-collapse models fail to describe the abundance ratios and large masses of iron-group elements. 
A model based on a 
multi-spot ignition of the
white dwarf can explain the observed high $M_{\rm Mn}/M_{\rm Cr}$ value (0.8--2.2).
The bar-like morphology is mainly due to a density enhancement in the center, 
given the good spatial correlation between gas density and X-ray brightness. 
The recombination ages and the Sedov age consistently
suggest a revised SNR age of  5--6~kyr. 
This study suggests that despite the presence of
candidate point sources projected within the boundary
of this SNR, W49B is likely a Type-Ia
SNR, which suggests that Type-Ia supernovae can
also result in mixed-morphology SNRs.
}


\titlerunning{Asymmetric Type-Ia supernova origin of W49B} 
\authorrunning{P. Zhou \& J. Vink}
\maketitle

\section{Introduction}

The study of supernova remnants (SNRs) provides information
about both the supernova explosions themselves, and about the environments
in which the supernova explosions took place. The environment often carries
 important information about the supernova progenitor itself,
such as whether
it formed in a star-forming region, 
and whether the progenitor shaped its own environment with a stellar wind.
In particular, massive stars are known to create large wind-blown bubbles
of several tens of parsecs in size \citep{weaver77,chevalier99}.

The morphology and spectra of  SNRs are determined by the
combined effects of both the intrinsic explosion properties and
the ambient medium in which they involve. Unfortunately, however, it is sometimes
difficult to disentangle the effects of explosion properties and the
environment in which they occurred. There are properties
that can be firmly attributed to the explosion properties, but also properties 
that may be attributed to
either the explosion characteristics or to the environment.
For example, for young SNRs it is clear that the abundance pattern provides
clear signatures of the type of explosion, with Type-Ia supernovae
producing more iron-group elements (IGEs), whereas core-collapse (CC) SNRs
are more abundant in oxygen, neon, and magnesium \citep{hughes95,vink12}.

Mixed-morphology SNRs are a special class of SNRs characterized by bright thermal
X-ray emission from their center, and a shell-type morphology
in the radio. 
Initially, it was noted that mixed-morphology SNRs show thermal
emission in the interior originating from low-abundant hot gas
\citep{rho98, jones98}. 
W49B was included in the list, but appeared to have
enhanced abundances. 
However, increasingly more of the originally mixed-morphology SNRs appeared to also have enhanced abundances in their interiors \citep{lazendic06b}. 
It is clear that  W49B, together with some other metal-rich cases
such as Sgr A East \citep{sakano04,park05} stand out.
In most reviews of mixed-morphology SNRs, W49B is listed as a mixed-morphology SNR \citep{lazendic06b, vink12, zhang15,dubner15}, and the definition of mixed-morphology SNR is in those cases based solely on different radio and X-ray morphology, and the fact that the X-ray emission is thermal in nature.
It is thought that mixed-morphology SNRs evolve in
denser environments, and since massive stars are associated with molecular
cloud environments, it is usually assumed that these SNRs are remnants
of CC supernovae. Indeed, some of the mixed-morphology
SNRs have associated young pulsars, proving that these SNRs are indeed
CC SNRs \citep[e.g., W44, IC~443,][]{wolszczan91, olbert01}.

Thermonuclear (or Type-Ia) supernova progenitors are
carbon-oxygen white dwarfs (WDs), which take a longer time to evolve ($>40$~Myr),
and, moreover, only explode if they accrete sufficient matter from a companion star 
\citep[the single-degenerate scenario;][]{whelan73}, or merge
with a companion carbon-oxygen WD \citep[the double-degenerate scenario;][]{webbink84}. By the time they explode, their
ambient medium does not necessarily contain any information anymore
about their progenitors.
The exact origin of
Type-Ia supernovae is still a source of debate 
\citep[see reviews][and references therein]{branch95,hillebrandt00,livio00,wang12,maoz14},
but also the manner in which the WDs
explode is uncertain, with  models involving deflagration \citep{nomoto84}, competing with so-called delayed detonation (DDT) models \citep{khokhlov91}. In general, Type-Ia SNRs are often to be found in less dense regions of the Galaxy. 
For example, SN\, 1006 is found high above the Galactic plane 
\citep[$b=14.6^\circ$, corresponding to $\sim 560$~pc at a distance of 
$2.18\pm0.08$ kpc,][]{winkler03}. 
The less disturbed
media in which they are  often found may account for the generally more symmetric morphology, as compared
to CC SNRs \citep{lopez11}.
On the other hand, the more symmetric morphologies of Type-Ia SNRs may also be caused by intrinsically more symmetric explosions.

The idea that Type-Ia progenitors do not shape the supernova environments has recently been challenged. For example,
it is clear that Kepler's SNR \citep[][for a review]{vink17a},
a Type-Ia SNR \citep{kinugasa99,reynolds94}, is evolving inside a bow-shock-shaped high-density region
caused by the wind from a progenitor system \citep{chiotellis12}. 
In contrast, the likely Type-Ia SNR RCW 86 
\citep[see][for a recent paper suggesting a core-collapse origin]{gvaramadze17}
seems to evolve inside the low-density environment created by a powerful low-density wind
\citep{williams11,broersen14}.
Type-Ia SNR Tycho is suggested to be overrunning a slowly expanding molecular 
bubble created by its progenitor's outflow \citep{zhou16a}. 
The middle-aged SNR G299.2$-$2.9 is a Type-Ia SNR showing asymmetries in the ejecta distribution due to an asymmetric explosion and/or a nonuniform surrounding medium \citep{post14, park07}.
Moreover, some Type-Ia supernovae may explode with intrinsic asymmetries \citep{ropke07, maeda10b},
which has been used to interpret the spectral evolution diversity observed in Type-Ia supernovae \citep{maeda10a}.

Many of the above-mentioned issues of relating SNRs and their environments to the explosion types,
come together in the peculiar SNR W49B,
which is suggested at a distance of 8--11.3 kpc \citep{radhakrishnan72, brogan01, chen14, zhu14}.
The X-ray emission from W49B is dominated by emission from
the center, which was initially attributed to the presence of a pulsar wind nebula \citep{pye84}, but
was not much later discredited by the fact that the X-ray spectra obtained by the EXOSAT
satellite displayed bright Fe-K lines \citep{smith85}.
Due to the centrally enhanced X-ray morphology, it was listed
in the first catalog of mixed-morphology SNRs, but with a metal-rich interior
\citep{rho98}. 
However, the brightness of the Fe-K lines
and detailed spectroscopy with ASCA \citep{hwang00} suggested that W49B is relatively young (1000-4000~yr)
compared to most mixed-morphology SNRs, although it does  share some characteristics with Sgr A, an East metal-rich,
mixed-morphology SNR \citep{maeda02}.
A property that W49B shares with many other mixed-morphology SNRs is that the plasma appears
over-ionized \citep{kawasaki05,yamaguchi09,miceli10}, rather than under-ionized, as in most SNRs. Like many mixed-morphology SNRs, \snr\ is a GeV gamma-ray source, but it is also a TeV gamma-ray,
which is more rare for this class of SNRs \citep{hess16}.

Although the X-ray spectrum of  W49B shows the SNR to be very iron-rich, it is usually assumed that
it is a CC SNR, like most mixed-morphology SNRs, albeit a peculiar one. 
\citet{hwang00} expressed some doubts, suggesting
that neither a CC origin, nor a Type-Ia origin could explain the measured abundances.
The brightness of the Fe-K lines, but also the peculiar, jet-like morphology of the ejecta, has been
interpreted as evidence that W49B is the result of a hypernova explosion \citep{keohane07}.
On the other hand, \citet{miceli06} compared the observed abundances with yields for hypernova and supernova nucleosynthesis 
and found better agreement for the abundances of \snr\ with models with a normal explosion energy ($10^{51}~\erg$).
More recently, \citet{lopez13a}, assuming W49B to be a CC SNR,
presented evidence that the supernova produced a black hole rather than a neutron star (NS).
The reason is that they did not find evidence for a cooling NS, similar to the X-ray
point source in Cas A \citep{tananbaum99}.

The study presented here was prompted by the many peculiarities of W49B. Most notably, we
were puzzled by the fact that black holes are thought to be the end products of the most massive
stars \citep[$> 25~\Msun$, e.g.,][]{heger03}, but
W49B seems to be evolving in a cavity of only $\sim 5$~pc radius \citep{keohane07}.
In contrast, a progenitor more massive than 25$\Msun$ is expected to create a cavity with a radius of at
least 20~pc \citep{chen13}.

With our study we therefore tried to investigate a) whether a cooling NS may, after all, be present,
given that W49B provides a spatially non-uniform X-ray background
that could hide a point source, and that the interstellar absorption
is relatively high ($N_\mathrm{H} > 10^{22}$~cm$^{-2}$; and b) whether W49B is indeed a CC SNR or even a jet-driven CC SNR, as often assumed.

To answer these questions we reanalyzed
the archival Chandra data, using a state-of-the-art adaptive binning method
for spatially resolved X-ray spectroscopy, and we made
a new search for X-ray point sources. We indeed found a few point sources, but our overall conclusion is that
the X-ray spectra fit better with a Type-Ia origin for W49B, and in particular that  the  abundance pattern best fits the multi-point ignition DDT models
of \citet{seitenzahl13a}.

\section{Data and analysis}

\subsection{Data}
\snr\ was observed with \Chandra\ in three epochs in 2000 (obs. ID: 117;  PI: Stephen Holt) 
and 2011 (obs. IDs: 13440 and 13441; PI: Laura Lopez), with exposures of 54, 
158, and 60 ks, respectively.
We retrieve three sets of Chandra data, 
which covered the SNR with the backside-illuminated S3 chip in 
the faint mode.
We use CIAO software (vers. 4.9 and CALDB vers. 4.7.7)\footnote{http://cxc.harvard.edu/ciao} to 
reduce the data, extract spectra, and 
detect the point-like sources.
Xspec (vers. 12.9.0u)\footnote{https://heasarc.gsfc.nasa.gov/xanadu/xspec}
is used for spectral analysis.

\subsection{Spatial-spectral analysis and adaptive binning method}

To optimize the binning of the X-ray data for spatially resolved study, we employ a 
state-of-the-art adaptive spatial binning method called the weighted Voronoi tessellations 
(WVT) binning algorithm \citep{diehl06},
which is a generalization of \citet{cappellari03} Voronoi binning algorithm. 
The X-ray events taken from the event file are adaptively binned with the WVT binning algorithm.

First, we use the observation with longest exposure (obs. ID: 13440) 
to generate the bins, with each bin containing about 3600 counts 
in 0.3--10 keV.
Among the 238 detected bins, 177 are located inside the SNR, with mean counts 
per pixel  ($1''$) larger than 5. 
Subsequently, we extract spectra of the 177 regions from
three Chandra data sets.
The combined data provide $\sim 6000$ counts in each spatial bin.
We then perform spectral analysis on the 177 bins associated with \snr\ 
by jointly fitting three groups of spectra from three observations.
For each spatial bin and each observation,
separate response matrix file and ancillary response file
are generated.
The background spectra are selected from a source-free region to the northeast of the SNR, which is at a similar Galactic latitude to the remnant.

Since previous studies \citep{kawasaki05, miceli06, miceli10, ozawa09, lopez13b, lopez13a} indicate
that the spectrum of W49B is best fit with either
a collisional ionization equilibrium (CIE) model or a recombining plasma model
(a non-equilibrium ionization model with over-ionization), or a combination of the two models, we started our
analysis by fitting each spatial bin twice, once with  a CIE and once with an over-ionization model;
in the end we selected the model that fits the spectra of a given spatial bin best with the smallest reduced chi-squared $\chi_\nu^2$; see Section~\ref{S:onecomp}).

\begin{figure*}
  \centering
  \includegraphics[angle=0, width=\textwidth]{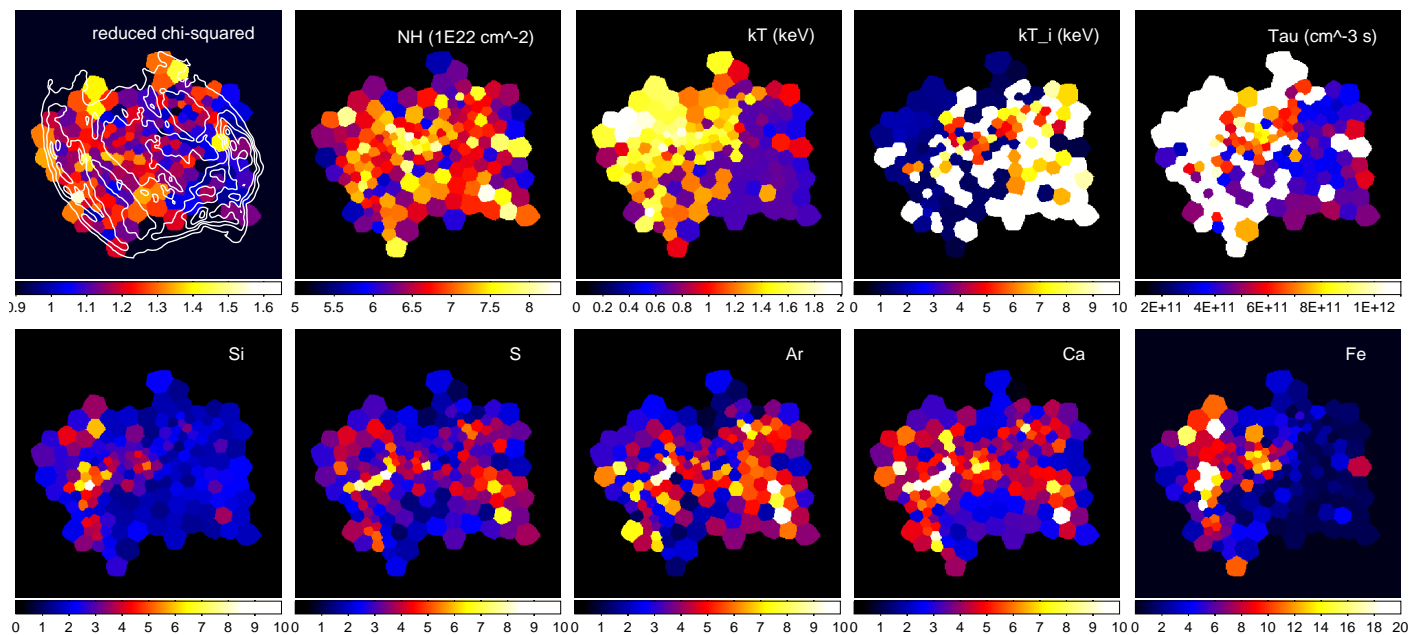}
\caption{Distribution of the parameters fitted with the best-fit single 
component model, 
which, for every cell, was taken to be either the absorbed "{\it vrnei}" or the absorbed "{\it apec}" model,
whichever produced the smallest $\chi_\nu^2$.
The first panel shows $\chi_\nu^2$ overlaid
with VLA 20 cm radio contours \citep{helfand06} in white color. }
\label{F:pars1}
\end{figure*}

\begin{figure*}
  \centering
  \includegraphics[angle=0, width=\textwidth]{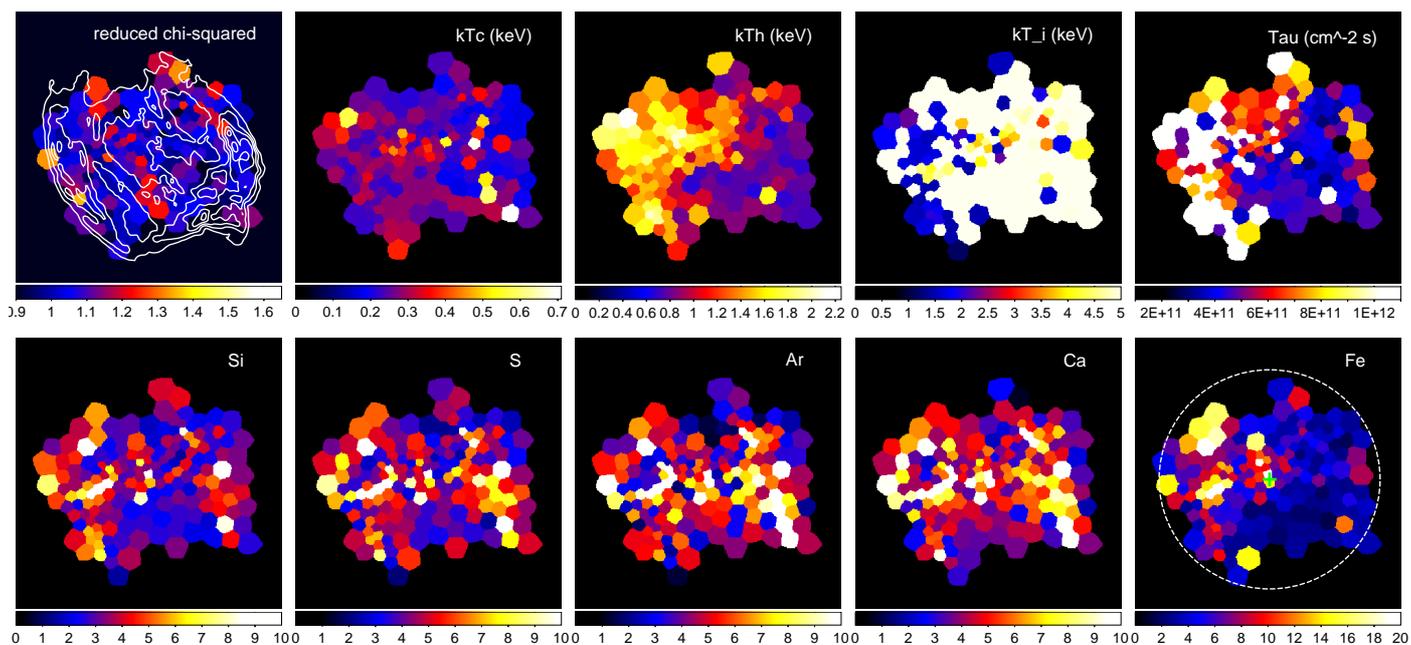}
\caption{Distribution of the parameters fitted with the best-fit double component model, 
which was either an absorbed "hot {\it vrnei} + cool {\it apec}" model, or an absorbed "hot {\it vapec} + cool {\it apec}" model.
The first panel shows $\chi_\nu^2$ overlaid with VLA 20 cm radio contours 
in white color. 
In the last panel, the dashed circle and the green cross sign denote the SNR sphere
used for density calculation and the sphere center used for abundance -- P.A. diagram in Figure~\ref{F:paabun}, respectively.
}
\label{F:pars2}
\end{figure*}

\begin{figure*}
  \includegraphics[angle=0, width=\textwidth]{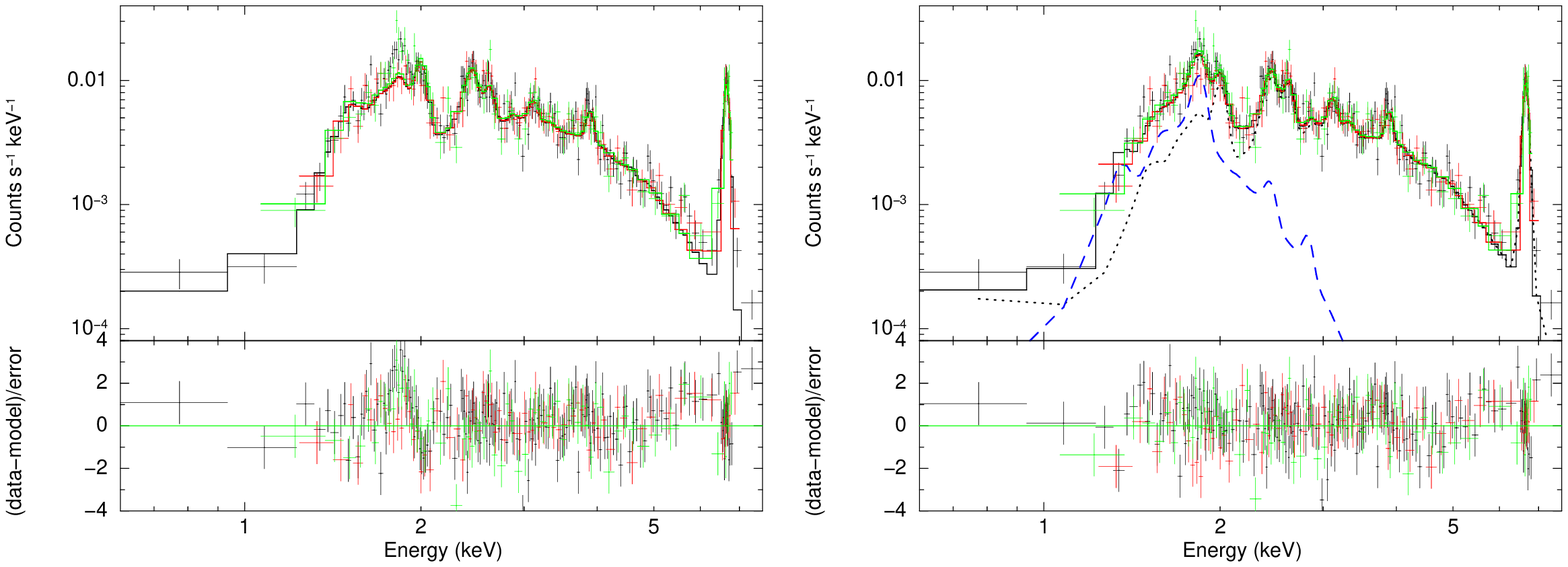}
\caption{Exemplified ACIS-S spectra from one bin in \snr\
fitted with one component ($vapec$; left
$\chi_\nu^2/d.o.f.=1.44/308$
) and two components 
($apec$ in blue dashed line plus $vapec$ in black dotted line; right; 
$\chi_\nu^2/d.o.f.=1.15/307$) models, 
respectively. 
The black, red, and green data correspond to the spectra from
the observations 13440, 13441 and 117, respectively.
}
\label{F:spec_27}
\end{figure*}

Because this single-component model did not
always give satisfactory fits, we decided to also use a two-component model,
combining a relatively cool CIE component with a hotter component, which could
be either a CIE or an over-ionized model (Section~\ref{S:twocomp}).
The two component model gives better fits to the spectra, but
a problem is that some of the parameters are correlated, so some additional
constraints had to be imposed.

One should be aware that the recombining plasma model, with two more
free parameters, may provide  slightly smaller $\chi_\nu^2$ than 
the CIE model even though sometimes  the two models are not statistically distinguished. 
Nevertheless,  we found that for the single-component case discussed in Section~\ref{S:onecomp}
the two models can be distinguished in 90\% of the extracted
regions (based on the F-test, 2$\sigma$ level), and
for the two-component case in Section~\ref{S:twocomp}, 
the two groups of models are distinguishable in 80\% of regions.

\subsection{Single thermal component} \label{S:onecomp}

We jointly fit the spectra in each bin with an absorbed CIE model ($vapec$) and 
an absorbed recombining plasma model ($vrnei$) in Xspec, and
select the model with smaller $\chisq$ as the best-fit model.
The two plasma models use the atomic data in the ATOMDB code\footnote{http://www.atomdb.org/}
version 3.0.7.
The Tuebingen-Boulder interstellar medium (ISM) absorption model $tbabs$ is used for calculation of the X-ray
absorption due to the gas-phase ISM, the grain-phase ISM, and the molecules in the ISM \citep{wilms00}. 
The $vrnei$ model describes a plasma that has cooled/heated 
rapidly from an initial 
temperature $kT_{\rm i}$ to a temperature of $kT$, whereas
the ionization state lags behind, and is characterized
by a recombination/ionization timescale $\tau_{\rm r}/\tau_{\rm i}$ 
applied to all ions.
When the ions are in the recombining (overionized) state, $kT_{\rm i}$
and the "ionization temperatures" of some ions $kT_z$ are 
larger than the current electron temperature $kT$. 
The lower limit on $\kTi$ is set to 2 keV to ensure that the recombining 
model is used, 
where the high ionization temperatures of Ar and Ca 
\citep[$kT_z=2.2$--2.7 keV;][]{kawasaki05} are also considered.
We set the upper limit of $kT_i$ to 10 keV, which is a very high value 
for the plasma in an evolved SNR.
We allow the abundances of Si, S, Ar, Ca, and Fe to vary 
and  tie the abundance of Ni to Fe. Lower-mass elements, such as O, Ne, and Mg, have X-ray line emission
below 1.8~keV, in the part of the spectrum that is heavily affected by interstellar absorption.
The abundances of these elements are therefore unconstrained by the fits, and we fixed
the abundances to their solar values.
The solar abundances of \citet{asplund09} are adopted in plasma 
emission and photoelectric absorption models.  
Compared to the older widely used abundances obtained by \citet{anders89},
the O and Fe abundances in \citet{asplund09} decreased by 42\% and 
32\%, respectively.
Therefore, adopting different solar abundances can result in 
differences to the obtained abundances (especially for Fe) and
absorption column density \footnote{For example, using the solar abundances of \citet{anders89} result in $\sim 30\%$ smaller $\NH$ and Fe 
abundance for the spectra showed in Figure~\ref{F:spec_27} (for both of the single and two component models).}.

The single thermal component model gives a $\chisq$ between 0.9 and 1.5 (degree of freedom (d.o.f.)=222--365; mean $\chisq$/d.o.f=1.16/291) across the remnant.
Figure~\ref{F:pars1} displays the spatial distribution of the best-fit parameters,
including the foreground absorption $\NH$, electron temperature $kT$, initial temperature $\kTi$
(equal to $kT$ when CIE is the best-fit model), recombination timescale $\tau_{\rm r}$ (equal
to the upper limit of $10^{14}\cm^{-3} \s$ when CIE is the best-fit model), and the abundances
of Si, S, Ar, Ca, and Fe.
The figure illustrates a large variation of the gas properties inside \snr. 
The $\kTi$ values can be constrained in only a small fraction of regions, 
while in the southwestern regions they run to the upper limit of 10 keV.

\subsection{Two thermal components} \label{S:twocomp}

Previous studies suggest the existence of a cooler ISM component \citep[$\sim 0.25$~keV;][]{kawasaki05}
in addition to the hot, metal-rich component \citep{hwang00, miceli06, lopez13b}.
Moreover, the infrared observations reveal that a large amount of dense 
gas ($\sim 500~\cm^{-3}$ and $\sim 800~\Msun$) is shocked by \snr\ \citep{zhu14}, 
which implies that the SNR in such an inhomogeneous environment should consist of 
more than one density/temperature component.
We therefore also apply a two-thermal-component model to fit the spectra, in an attempt
to check  whether or not a cool component is needed across the SNR.
The colder $apec$ model with solar abundances is added to the absorbed 
$vrnei$/$vapec$ model and then used to fit the 177 groups of spectra.
The soft X-ray photons $\lesssim 2$ keV (dominated by the cooler component) 
suffer heavy absorption, which results in
some degeneracy between the cool component
temperature $\kTc$ and $\NH$ in the spectral fit.
We therefore assume that the foreground absorption of \snr\ is not changed across the small angular extent of the SNR 
 ($<5'$) and fix the $\NH$ value to $8\E{22} \cm^{-2}$. 
This value is close to the best-fit $\NH$ for the global spectra, and is also similar to the 
mean $\NH$ value if $\NH$ is allowed to vary for the 177 spectra.
The upper limit for $\kTc$ is set to 0.7~keV to ensure that it is smaller than $\kTh$. 

\begin{figure*}
  \centerline{
  \includegraphics[angle=0, width=\textwidth]{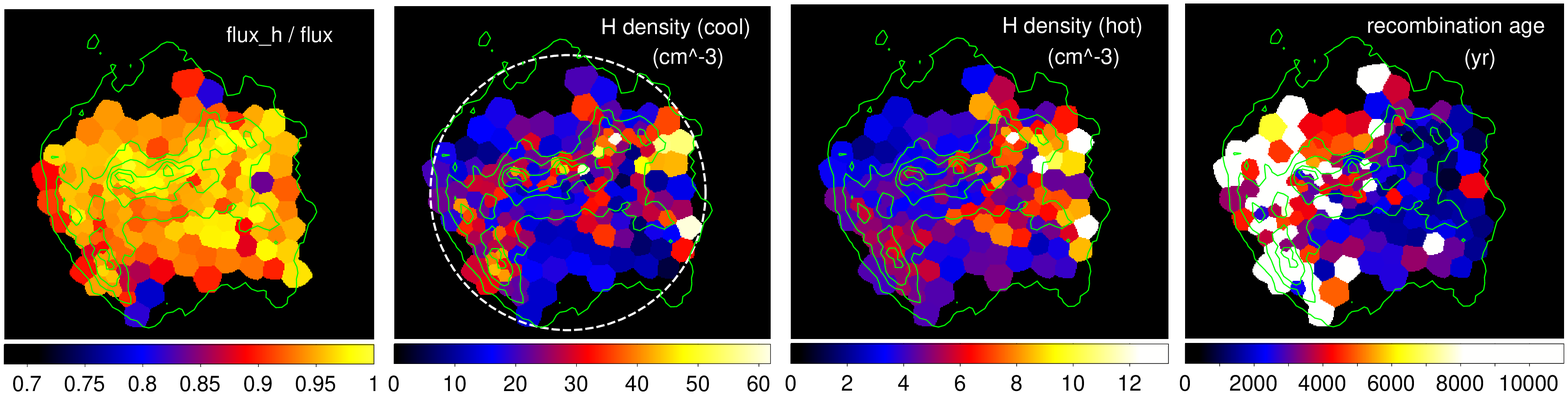}
   }
\caption{Distribution of the hot-to-total flux ratio (0.75--0.99), hydrogen
densities in the cold and hot phases, respectively, and recombination age 
of the hot components, overlaid with green contours the Chandra X-ray emission in 0.3--10 keV. The dashed circle denotes the SNR sphere for the density calculation.}
\label{F:density}
\end{figure*}

Figure~\ref{F:pars2} shows the detailed distribution of the gas 
properties in \snr\ based on the two-component model,  
including $\chisq$, the temperatures of the colder and hotter component 
($\kTc$ and $\kTh$, respectively), $\kTi$, $\tau_{\rm r}$, and the abundances 
of Si, S, Ar, Ca, Fe elements.
The Si distribution is smoother in the two-thermal-component model.
Here the upper limit of $\kTi$ is set to 5 keV, which appears to be a limit
to the electron temperatures found in young SNRs \citep{vink12}.
We note that the proton temperature may be higher, but the ionization properties
of the plasma are determined by the electron temperatures.
Moreover, the adopted upper limit value is indicated by the best-fit 
$\kTi$ values ($\sim 5$~keV) in the SNR center.

Although the overall temperature and abundance patterns obtained with
single and double component models appear to be similar,
the two-thermal-component model better describes the spectra than the  model with only one thermal component: it results in smaller $\chisq$ values (mean value of 1.06
and d.o.f of 290),
as well 
as smaller residuals in the spectral fit for the Si-XIII line emission
for the range 1.8--2.2 keV (Fig.~\ref{F:spec_27}).
The cooler component mainly affects the soft spectra $\lesssim 2$ keV 
and the best-fit values of Si 
(see Figure~\ref{F:spec_27} for an example of the spectra fitted with 
one component and two components models, respectively).
It is statistically meaningful to add the extra thermal component to the 
single-thermal-component model to improve the spectral fit of majority regions
according to an F-test analysis
(mean null hypothesis probability $10^{-7}$, adopting a mean d.o.f.\ of 291 
and $\chi_\nu^2=1.16$  in the single thermal model).
For 92\% of the bins, the two-component model significantly improves the fit with an 
F-test probability less than the typical value of 0.05 ($2\sigma$
level), while
for the other 8\% bins near the western boundary the improvement is less significant.

The X-ray flux of \snr\ is dominated by the hot component, as
indicated by the large fraction of the hot-component flux to the total 
flux in 0.5--8 keV (flux\_h/flux=0.78 --  0.99; see the left panel of Figure~\ref{F:density}).
The X-ray contribution from the cool component (1-flux\_h/flux) is larger 
near the SNR shell than in the SNR interior. 
As a result, the single component fit with $\NH$ free gives
smaller $\NH$ values near the SNR edge to compensate the flux of 
the missing cool component (see the $\NH$ distribution in Figure~\ref{F:pars1}).
Therefore, the variation of $\NH$ is likely to be much smaller than
suggested by the single-component model.

A temperature gradient for the hot component gas ($\kTh$) is revealed with an orientation
 northeast to southwest (see Figure~\ref{F:pars2}).
The $\kTh$ is as high as $\sim 2.2$ keV in the northeast but decreased to $\sim 0.7$~keV 
in the southwest.
The $\kTc$ shows some variation across the SNR with a mean value of 0.27 keV. 
In a small fraction of regions with $\NH$ deviated largely from 
$8\E{22}\cm^{-2}$, $\kTc$ can be affected by $\NH$ due to the degeneracy between them. 
For example, a region in the southwest shows large $\kTc=0.7$ keV, which could be reduced to 0.18~keV if $\NH$ is free ($\sim 10^{23}\cm^{-2}$).

The recombining plasma appears to be present throughout the SNR, except in the southeast shell, and
occupies about three quarters of the studied area.
The  recombination timescale varies in the range 1--$10\E{11} \cm^{-3}\s$.
The recombining gas and CIE gas can be distinguished by comparing its $\kTi$ and 
$\kTh$ or by checking $\tau_{\rm r}$ values: the recombining gas shows $\kTi > \kTh$ and 
$\tau_{\rm r} \lesssim 10^{12} \cm^{-3} \s$.
The CIE region is colored in white in the $\tau_{\rm r}$ panel of Figure~\ref{F:pars2}
(upper-right).
Our study confirms the existence of recombining plasma in the center and west as pointed out 
by previous studies \citep[e.g.,][]{miceli10, lopez13b}.
We note that the recombining plasma is also patchily distributed in the northern hemisphere of the remnant, but here
the recombination timescales are generally much longer than in the southwest.

\subsection{Density and mass of the shocked gas}

\snr\ has a centrally bright X-ray morphology, indicating an enhanced
plasma density in the interior, unlike the shell-type
SNRs in Sedov phase with a density enhancement at the shells
\citep{borkowski01}.
Since the SNR is highly structured without a clear understanding
of its three-dimensional density distribution, 
any complicate hypothesis on the gas distribution (e.g., bar+shell, disk+shell) 
could differ from the real distribution and introduce unknown uncertainties.
Therefore, we take the simplest assumption that the plasma is uniform in 
each bin, so as to provide a mean gas density of the X-ray-emitting gas.
We note that the following results are based on this oversimplified assumption of the geometry.

We estimate the mean density $\nH$ for a given bin using the normalization parameter in Xspec ($norm=10^{-14}/(4\pi d^2) \int n_e n_H dV$, where $d$ is the distance, $n_e$ and $n_H$ are the electron and H densities in the volume $V$; $n_e$=1.2$n_H$ for fully ionized plasma with solar or enhanced metal abundances in \snr)
and an assumed prism geometry 
for each bin.
Each prism has an area of the region and a depth across the SNR $l(r)=2\sqrt{\Rs^2-r^2}$, 
where the radius of the SNR is $\Rs=2\farcm{2}$, 
corresponding to 6.4~pc at a distance of 10 kpc,
and $r$ is its projection distance to the assumed SNR center.
The SNR circle denoted in Figure~\ref{F:density} encloses
all regions, except a bin in the southwest, where
its density and mass are not calculated.
The two-temperature gas is assumed to fill the whole volume ($f_{\rm c}+f_{\rm h}=1$)
and in pressure balance ($n_{\rm c} T_{\rm c}= n_{\rm h} T_{\rm h}$),
where $f$ and $n$ are the filling factor and hydrogen density, respectively, 
and the subscripts ``c'' and ``h'' denote the parameters for the cool and hot
phases, respectively.

The distributions of the densities are displayed in Figure~\ref{F:density}.
The brightness of the X-ray emission in 0.3--10 keV is overlaid with contours
for comparison purposes.
The hydrogen density $n_{\rm c}$ in the cooler component gas is clearly enhanced along
the ``bar-like'' feature across the SNR center and the shells in the eastern and 
western sides; it displays a good spatial correlation with the X-ray brightness.
An enhancement of the density in those regions is 
also present in the hot component.
The mean hydrogen density of the colder and hotter components are $24 \cm^{-3}$
and $5 \cm^{-3}$, respectively, and the mean filling factor of the hotter
phase is 40\%.
The total masses of the X-ray-emitting gas are $M_{\rm c}= 484_{-34}^{+41} \du^{2.5} \Msun$ 
for the cool phase and $M_{\rm h} = 52\pm8 \du^{2.5} \Msun$ for the hot phase, 
where $\du = d/(10~{\rm kpc})$ is the distance scaled to 10~kpc.

The mass uncertainties are calculated by using the 90\% uncertainties 
of the normalization parameters and the filling factors 
of the 176 bins, where the uncertainties in the filling factors are
incorporated into 
the overall error estimates,  also taking into account uncertainties in the temperatures and normalizations of the two components. 
If the errors are asymmetric, we take the largest
error. 
The systematic uncertainty is dominated by assumptions about  the volume $V$ or the depth 
$l$ of the X-ray emitting gas, since the gas mass in bin $i$ weakly depends on 
them ($M(i)\propto V(i)^{1/2}\propto l(i)^{1/2}$ with given distance and $norm$). 
Since mixed-morphology SNRs generally have a relatively smooth radial 
density distribution, we assume the X-ray-emitting gas fills the whole SNR from front to back.
This introduces some uncertainty, as the real depth, $l(i)$, can be smaller.
In the extreme case that the X-ray emission is only arising from a thin shell with 
a thickness of $1/12 R_{\rm S}$,\footnote{For a uniform density and a shock compression ratio
of four, mass conservation suggests that for shell-type SNR the shell should
have a thickness of
 approximately one twelfth $R$: $4\pi R^2\Delta R (4\rho_0)=4\pi/3 R^3 \rho_0$.}  the masses are $202^{+18}_{-14}~\Msun$ 
and $21\pm 3\Msun$ 
in the cool and hot components, respectively, which puts very conservative 
lower limits on  the gas masses in \snr.
We note that the thin shell geometry does not agree with the overall, center-filled, X-ray morphology of
\snr.

$M_{\rm c}$ is much larger than what could have been produced by the progenitor wind 
or the ejecta.
Hence, the cool component is dominated by the heated interstellar gas. 
The mass of the hot component is also too large to correspond to the total supernova ejecta mass, which 
suggests that it consists of a mixture
of ejecta and circumstellar material. 
An alternative could be that the hot component consists of almost
pure ejecta material, with almost no hydrogen. In that case, the bremsstrahlung continuum is
also dominated by electron-metal interactions rather than electron-proton interactions.
The high average charge of the atoms and the enhanced ratio of electrons to atom results then in a higher emissivity per atom, which
results in a lower mass for the hot component \citep[][section 6.1 and 10.3]{vink12}.
Abundance ratios are not very sensitive to whether the gas is metal-pure, or hydrogen-rich with enhanced
metal abundances.
Currently with CCD X-ray spectroscopy we cannot distinguish
between a pure metal plasma and a hydrogen-rich  plasma with enhanced abundances.
However, in the future, high-resolution imaging spectroscopy with the XARM (X-ray Astronomy Recovery Mission)
and/or
Athena could perhaps distinguish these two cases spectroscopically. Given the age of \snr\ (see also below)
a mixture of hydrogen-rich and metal-rich ejecta seems the most likely scenario.

\subsection{Abundance and distribution of the ejecta} \label{S:ejecta}

\begin{figure}
  \centering 
  \includegraphics[angle=0, width=0.49\textwidth]{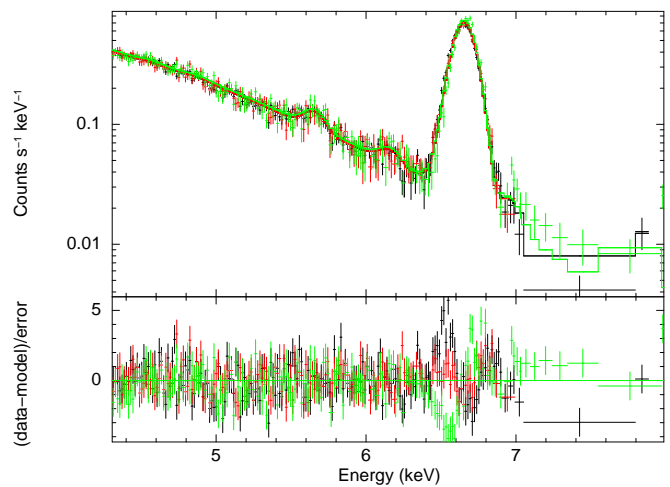}
\caption{
The global spectra of \snr\ in the 4.3--8.0 keV range fitted with a recombining 
plasma model $vvrnei$. 
The black, red, and green lines correspond to the spectra from
observations 13440, 13441 and 117, respectively.
}
\label{F:globalspec}
\end{figure}

The abundances and distribution of heavy elements have played an important role in
probing the explosion mechanism of SNRs.
We obtained average abundances 
[Si]=$3.4^{+0.8}_{-0.7}$, [S]=$4.9_{-1.0}^{+1.1}$, 
[Ar]=$5.1_{-1.0}^{+1.2}$, [Ca]=$5.2_{-1.0}^{+1.2}$, and [Fe]=$5.7_{-1.1}^{+1.2}$.
The average abundance values here are based on the weighted sum  
(by the estimated gas mass for each sub-region) over all individual sub-regions,
and are insensitive to the emission volume assumption as long as the emission volume 
does not vary sharply at different regions.
The abundance of Fe is greatly enhanced in the east of the SNR,
while the Si, S, Ar, Ca elements are highly enhanced across the SNR,
especially along the east and west (nearly axial symmetric distribution; see Section~\ref{S:aspherical}).

\snr\ is the first cosmic source in which Cr and possibly Mn emission were
found \citep{hwang00}.
The global spectra of \snr\ in 4.3--8.0 keV show a clear
Cr line at $\sim 5.6$~keV and a Mn bump at $\sim 6.1$~keV 
(see Figure~\ref{F:globalspec}). 
We fit the global spectra with the $vvrnei$ model
and obtained [Cr]=$6.6\pm 0.8$, [Mn]$=12.5\pm2.7$
(see Figure~\ref{F:globalspec}; $\chi_\nu^2/d.o.f=1.74/527$; 
$kT=1.48^{+0.07}_{-0.01}$~keV,  $\NH$ is fixed to $8\E{22}\cm^{-2}$, 
$kT_{\rm i} \ge 4.6$ 
keV, [Fe]=$3.2\pm0.1$;  [Ni] cannot be constrained and thus tied to [Fe]; 
$\tau_{\rm r}=6.0^{+1.1}_{-0.6}\E{11}~\cm^{-3}\s$).
The thermal plasma model constrains the Mn K line flux to $1.6\E{-5} \cm^{-2}\ps$,
similar to that obtained by \citet{miceli06} and \citet{yang13}.
The CIE model gives a slightly larger $\chi_\nu^2$ (1.78), 
with [Cr]=$5.9\pm0.7$,
[Mn]$=13.9^{+4.3}_{-4.1}$.
Hereafter, we use the best-fit abundance results of the $vvrnei$ model.
The residuals at around 6.6~keV are likely due to the gain shifts
for ACIS, but the shifts are within typical 0.3\% systematic uncertainties 
in the ACIS gain (response from the CXC  calibration scientists)\footnote{http://cxc.harvard.edu/cal/summary/Calibration\_Status\_Report.html}.
The Fe abundance obtained by fitting the global spectra is smaller
than the mass-weighted abundance ($5.7^{+1.2}_{-1.1}$), since the Fe 
element and the plasma properties vary across the SNR (see Figure~\ref{F:pars2}).

\begin{figure}[tbh!]
\centering
\includegraphics[angle=0, width=0.5\textwidth]{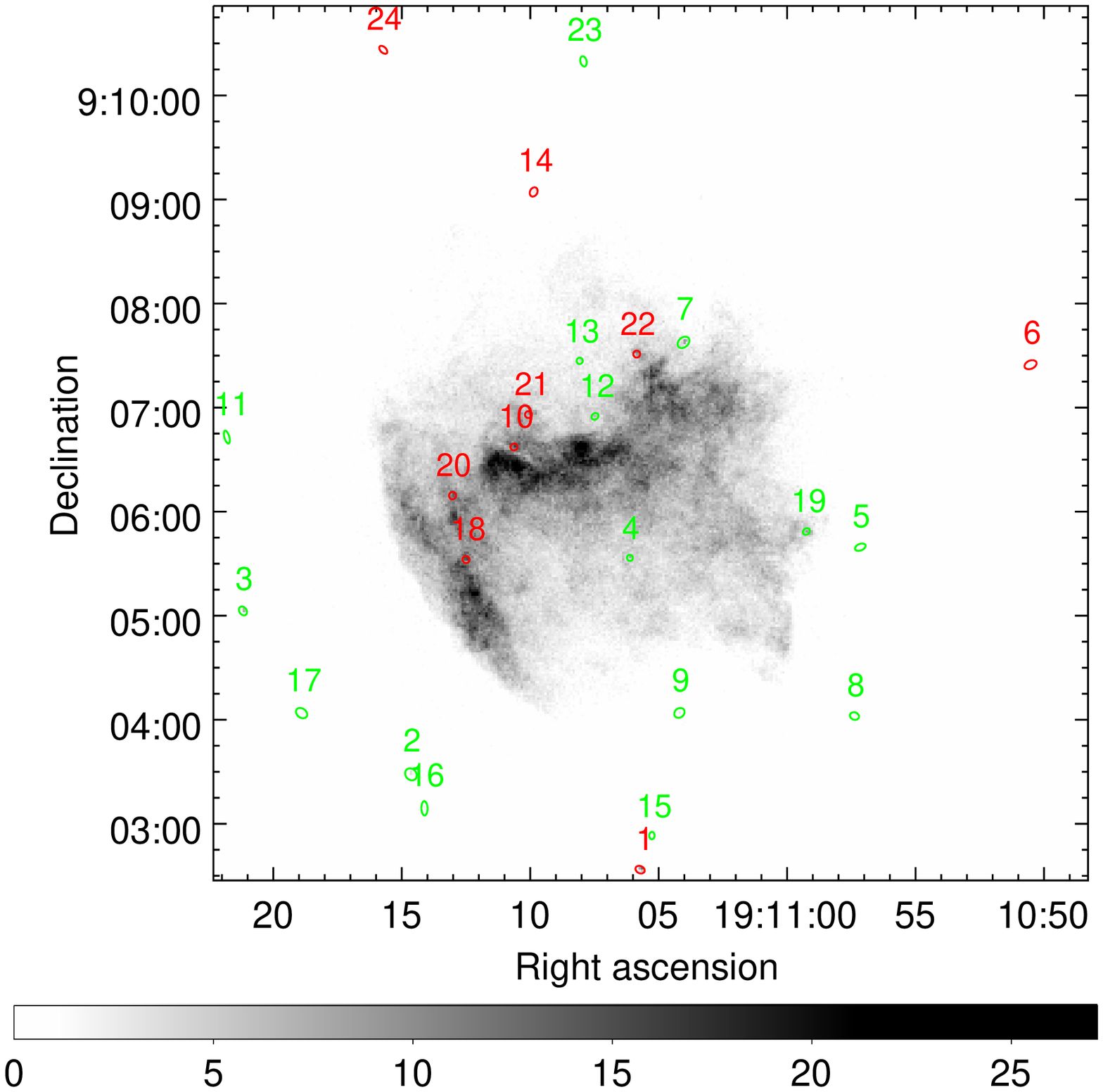}
\caption{The detected point-like sources in the \Chandra\ 0.7--5 keV raw image of \snr.
  The color-bar shows the scale of the counts per pixel ($0\farcs{5}$).
  The details of the sources in the ellipses are summarized in Table~\ref{T:ps}.
  The red regions denote the sources with best-fit blackbody luminosity in 
  the range $7.0\E{32}$--$2.7\E{34} \du^2 \erg\ps$ 
  corresponding to the luminosity
  range of NSs at the age of 5--6~kyr predicted with the minimal cooling paradigm
  (see the Appendix and Figure~\ref{F:coolcurve} for details).}
\label{F:ps}
\end{figure}
\begin{figure}
  \centering 
  \includegraphics[angle=0, width=0.47\textwidth]{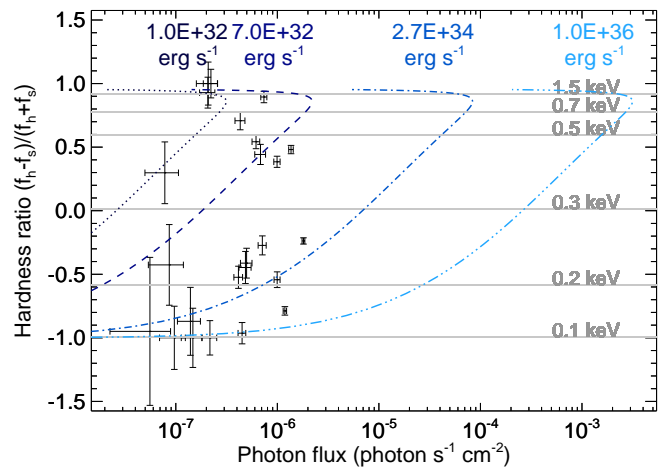}
\caption{
The hardness ratios (see the definition in Appendix A) and photon fluxes of the 24 point-like sources as a function
of the blackbody temperature and luminosity at a distance of 10~kpc.
The curved lines from the left to right indicate the luminosity from $10^{32} \erg\ps$
to $10^{36}\erg\ps$.
An absorption of $\NH=8\E{22}\cm^{-2}$ is assumed for all the sources.
}
\label{F:hr_8e22}
\end{figure}

\subsection{Point-like sources in the vicinity}

\snr\ has been considered to host a black hole, 
as no potential NS was detected
down to an X-ray luminosity of $2.7\E{31}\erg\ps$, and a core
collapse origin was assumed \citep{lopez13a}. 
However, the remnant reveals a very nonuniform X-ray brightness, which provides a spatially 
varied background for point-source detection. 
After considering the bright background from the SNR plasma, we detected 
24 point-like  sources in, or in the vicinity of, \snr\ using the Mexican-Hat 
wavelet detection method ($wavedetect$ in CIAO; see Figure~\ref{F:ps}). 
The point spread function (PSF) of the chip and the vignetting effect are taken
into account.
The applied significance threshold of source identification ($10^{-6}$) corresponds 
to one false detection in the image.

We derived the blackbody temperatures and luminosities of the sources
based the photon fluxes $f$ and hardness ratios (see Figure~\ref{F:hr_8e22}).
The luminosities of these sources are in the range $10^{32.1-36.2}~\erg\ps$
at an assumed distance of 10~kpc.
The details of the detection method and analysis are elaborated in Appendix A.
Detailed information about these sources is summarized in Table~\ref{T:ps}.

\section{Discussion}

Here we revisit the progenitor problems and discuss the CC (normal 
and energetic) and Type-Ia scenarios mainly 
based on three properties: 
metal abundances, metal distributions, and environment.
The metal abundances can be  compared to predictions of
supernova nucleosynthesis models, the distribution of metals reveals explosion (a)symmetries, and the density distribution provides information on the circumstellar environment.
We will also examine the SNR properties and discuss the origin of
the bar-like morphology according to the spatially resolved
spectroscopic analysis.

\subsection{CC scenario and its problems}

\subsubsection{Cavity} 

\snr\ is suggested to be interacting with a molecular cavity surrounding the
remnant \citep{chen14, zhu14}. 
A massive star can evacuate a hot, low-density bubble with its strong stellar wind
during the main sequence stage. 
The slow and dense wind during its later red supergiant stage will, however,  in most cases not 
reach the main sequence cavity shell.
In the molecular environment with a typical pressure of $p/k\sim 10^5 \cm^{-3} \K$
\citep{chevalier99, blitz93}, the maximum radius of the bubble $R_b$ is determined by 
the progenitor mass $M_*$: $R_b\approx1.22 M_*/\Msun -9.16$~pc \citep{chen13}.
This linear relation is valid for stars with masses less than $25\Msun$, while beyond
the mass the bubble could be larger due to the contribution from the fast wind 
in Wolf-Rayet stage.
If the molecular cavity surrounding \snr\ was created by the main sequence
wind of a massive progenitor, a bubble with a radius of $6\farcm{4}$ would suggest 
a progenitor mass of $\sim 13\Msun $ 
\citep[see also][]{chen14, zhang15}.
A smaller bubble size ($\sim 5$~pc) is also likely as suggested by \citet{keohane07}, and indicates a progenitor mass 
$\lesssim 13 \Msun$.
This is inconsistent with the idea that the supernova explosion resulted in the creation of a black hole \citep{lopez13b}, which is considered to require a $>25\Msun$ progenitor. A $>25\Msun$ progenitor instead would create
a wind  bubble in the molecular cloud of $\sim 21$~pc.

\subsubsection{Missing compact object?} 

\begin{figure}
  \centering 
  \includegraphics[angle=0, width=0.48\textwidth]{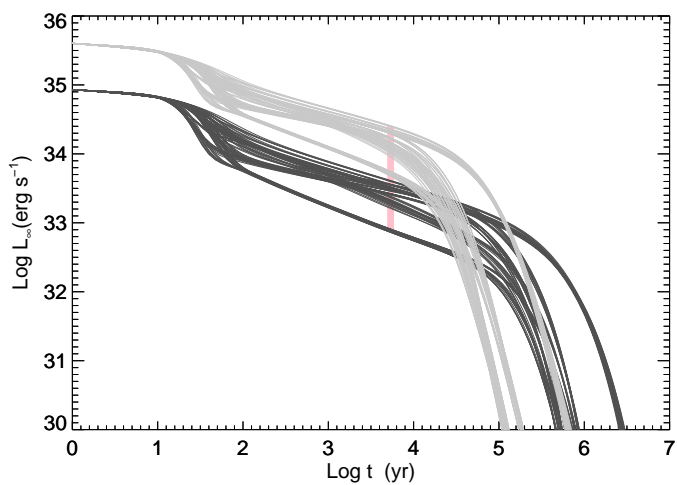}
\caption{
The predictions of NS cooling according to the minimal cooling paradigm \citep[see][]{page09}.
The light- and dark-gray lines represent the cooling curves of NSs
with light- and heavy-element envelopes, respectively.
The curves for intermediate-element envelopes are between them.
Each envelope group includes 100 curves corresponding to four choices of
$^3P_2$ gaps, and five choices of neutron $^1S_0$ and proton
$^1S_0$ gaps. 
The pink belt indicates the luminosity range of an NS
in an age range of 5--6 kyr.
}
\label{F:coolcurve}
\end{figure}

A massive single star with $M_*\lesssim~20$--$25\Msun$ ends its life with 
a NS \citep{heger03}.
The NSs, which are born extremely hot ($>10^{10}$~K), cool predominantly via 
neutrino emission from the interior for $\sim 10^4$--$10^5$ years after birth.
The cooling curve or luminosity-age relation depends sensitively on the equation 
of state of dense matter, NS mass, and envelope composition.
\citet{page04, page09} proposed the minimal cooling scenario including Cooper pair
breaking and formation process to explain the observed luminosities of considerably
young NSs.
According to the minimal cooling scenario, young NSs radiate 
X-ray emission in the initial few kyrs and their temperatures and
luminosities are a function of the NS age.

Figure~\ref{F:coolcurve} shows the cooling curve (luminosity evolution)
of NSs based on the minimal cooling paradigm\footnote{cooling code from Dany
P. Page: http://www.astroscu.unam.mx/neutrones/NSCool/}, which assumes that 
no enhanced neutrino emission is allowed in NSs \citep[see][]{page09}.
All models are for 1.4~$\Msun$ stars built using the NS equation of state of \citet{akmal98}.
The model predicts a luminosity range of $L_\infty=
7.0\E{32}$--$2.7\E{34} \erg\ps$ for the NS at an age of \snr\ (5--6~kyr; see discussion in
Section~\ref{S:age}).

Nine of the 24 detected point-like sources are in the luminosity range 
predicted for an NS at an age of 5--6~kyr according to the minimal cooling paradigm (regions 
labeled in red in Figure~\ref{F:ps}; distance of 10~kpc is assumed;
between the dashed and dot-dashed curves in Figure~\ref{F:hr_8e22}).
We note that some of the sources may not be real point sources,
but compact clumps of gas associated with \snr, while sources
outside the SNR boundary would need a large transverse velocity ($\gtrsim 10^3~\km\ps$) to establish a connection with the SNR.
As all the point sources are off-center,
the NS may have been born with a high velocity ($>400~\km\ps$ assuming an SNR age of 6~kyr)
if \snr\ was caused in a CC explosion.
Therefore, even if \snr\ is a CC SNR,  there is still
the possibility that it contains a cooling NS,
rather than a black hole, as suggested by \citet{lopez13a}.

\begin{figure*}
  \centerline{
  \includegraphics[angle=0, width=\textwidth]{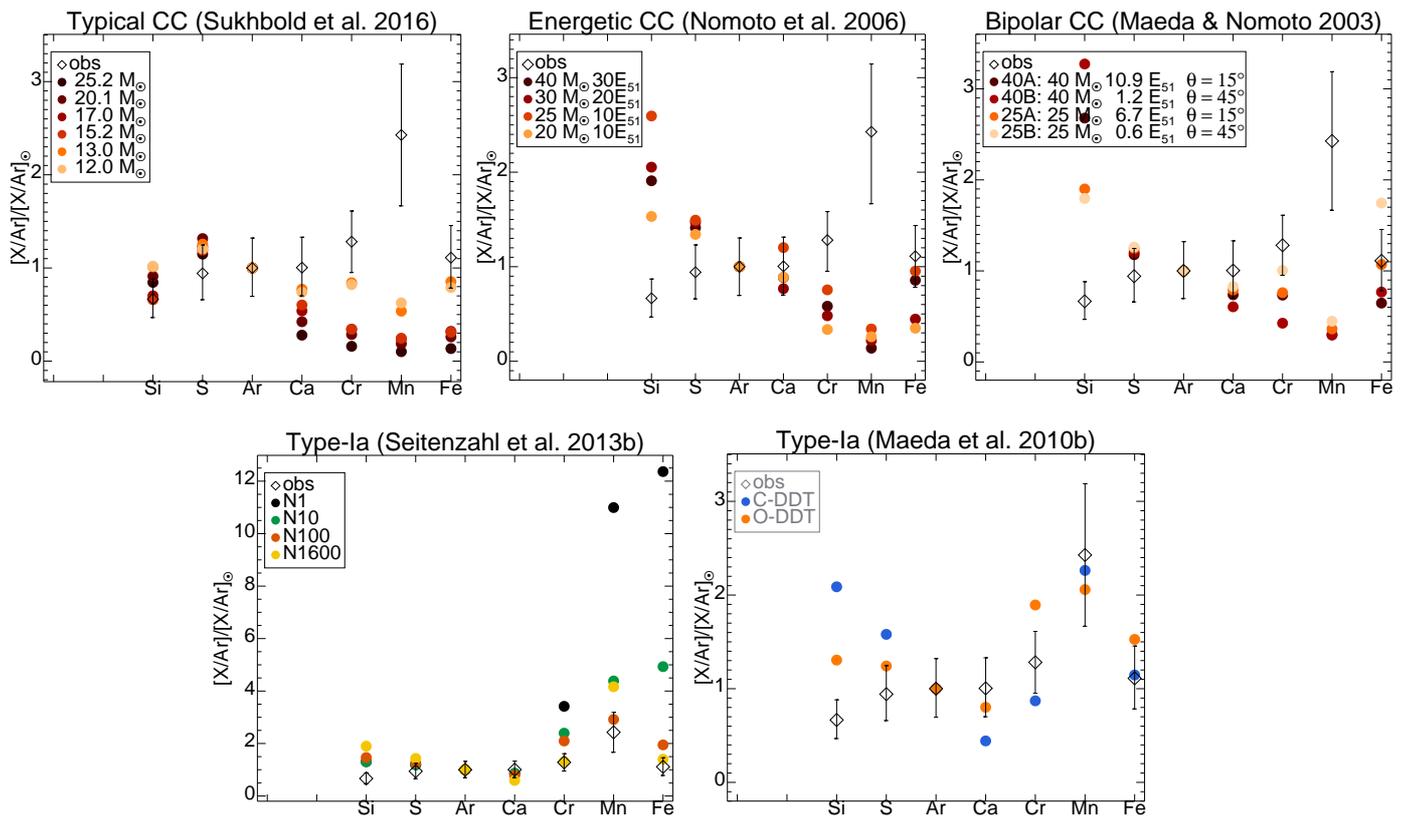}
  }
\caption{
The abundances of Si, Ar, Ca, Cr, Mn, and Fe relative Ar (with error bars) compared with 
the predictions of supernova nucleosynthesis models.
The upper panels show the CC spherical explosion models (left), and
energetic spherical (middle) and bipolar explosion models (right; the explosion energy are 10--30 $\times
10^{51} \erg$, $E_{51}$ stands for $10^{51} \erg$, $\theta$ is the
opening half-angle of the jet) for different progenitor masses.
Bottom-left panel shows three-dimensional DDT Type-Ia models, where the deflagration
is ignited in 1, 10, 100, and 1600 spherical sparks, respectively, near the WD center.
Bottom-right panel compares the DDT models which followed the
two-dimensional spherical deflagration (C-DDT) and extremely off-center deflagration
(O-DDT) ignitions,  respectively.
Solar progenitor metallicity is used for all models.
}
\label{F:nucleo}
\end{figure*}

\subsubsection{Metal abundances and yields} \label{S:CC_abun}

For a CC SNR showing super-solar abundances, the progenitor mass could
be estimated by comparing the abundance ratios with those predicted by the nucleosynthesis
models, since the nucleosynthesis yields of CC SNe are related to the progenitor mass,
stellar metallicity (solar assumed in this study), and explosion energy \citep[e.g.,][]{nomoto06}.
Figure~\ref{F:nucleo} shows the comparison of the abundances of Si, Ar, Ca, Cr, Mn, and Fe relative to Ar with the predictions of the spherical supernova explosion model  \citep{sukhbold16}, hypernova model \citep{nomoto06}, and bipolar explosion model \citep{maeda03}.
The observed abundance ratios are ascending with the atomic weight from Si to Mn.
We found that none of the available nucleosynthesis models explain the ascending abundance ratios.
The less massive stars (12--13~$\Msun$) provide a flatter abundance pattern as function 
of element mass, whereas the very 
massive progenitors and energetic models produce sharper descending abundance ratio patterns
(less IGEs), which are more deviated from the observed ratios.

Besides the inconsistent abundance ratios, none of the above-mentioned CC models 
explain the observed amount of IGEs.
The Cr, Mn, and Fe ejecta masses in the hotter-phase gas  are 
$M_{\rm Cr}=4.9\pm 1.1 \E{-3}~\Msun$, 
$M_{\rm Mn}=6.6\pm 1.9\E{-3}~\Msun$, 
$M_{\rm Fe}=0.32^{+0.10}_{-0.09}~\Msun$, 
respectively.
The nucleosynthesis models used in this study predict higher iron-group 
production for the stars with higher masses.
A $25.2~\Msun$ star produces $6.4\E{-4}~\Msun$ of Cr,
$3.2\E{-4}~\Msun$ of Mn, and $0.048~\Msun$ of Fe 
in a spherical explosion, each of which is about an order 
of magnitude lower than 
the value obtained from X-ray observations.
A $25~\Msun$ undergoing energetic ($10^{52} \erg$) 
spherical explosion 
produces slightly more Cr, Mn, and Fe 
($0.001~\Msun$, $3.5\E{-4}~\Msun$, and 
$0.007~\Msun$, respectively) compared to  
normal explosion energies, but this remains insufficient to create the observed yields.
Similarly, all of the bipolar CC explosion models predict 
much smaller IGE yields than the observed values.
The bipolar energetic explosion of a $25~\Msun$ star 
(model 25A) produces 
Cr, Mn and Fe masses of $7.5\E{-4}~\Msun$, $2.3\E{-4}~\Msun$,
and $0.082~\Msun$, respectively.

Hence, the CC models fail to explain the observed abundance ratios and 
under-predict the mass of IGEs.
Due to the large abundances of Cr and Mn, this conclusion is unchanged,
even taking into account the large systematic uncertainties 
in $M_{\rm h}$ ($>18~\Msun$; 
see Section~\ref{S:twocomp}).

\subsection{Aspherical Type-Ia explosion} \label{S:ia}

\subsubsection{Metal abundances and yields}

Type-Ia SNe are the dominant factories of IGEs \citep[see a recent review of][]{seitenzahl17}.
The large masses of Cr, Mn, and Fe and the high IGE/intermediate-mass element (IME) ratio in \snr\ 
clearly suggest a Type-Ia origin.
As shown in the bottom panels of Figure~\ref{F:nucleo}, we compare 
the abundances ratios of the SNR to the predicted results of different 
Type-Ia SNe models, including DDT models followed with spherical or 
extremely off-center slow flagrations, according to the 
three-dimensional model from \citet{seitenzahl13a} and two-dimensional models from \citet{maeda10b}.
The classical deflagration model W7 \citet{nomoto84} was excluded since it
over-predicts the Fe abundance.

For the models in the bottom-left panel of Figure~\ref{F:nucleo}, the 
deflagration is assumed to be ignited from different numbers 
of sparks in a WD with a central density of $2.9\E{9} \g\cm^{-3}$.
The model N1 is for the single spot ignition and N100 denotes the 
100-spot ignition. 
The occurrence of multi-spot ignitions, which covers  a range of offset radii, is a 
probable consequence of the turbulent convection of the WD prior to 
the thermonuclear runaway \citep{garcia-senz95, 
woosley04, iapichino06}.
The fewer sparks burn less materials to power the expansion of the WD
and thus the deflagration is weaker. The moderately strong expansion 
causes a high central density at the onset of detonation, and most of the 
remaining fuel is burned to IGEs by detonation. Conversely, a larger
number of ignition kernels produces relatively more IMEs during  incomplete 
burning \citep{seitenzahl13a}.

The abundance ratios in \snr\ can be well described by the multi-spark ignition models N100 and N1600, except for the
element Si.
Hence, the model with moderate to large numbers of ignition sparks
reproduces the abundances and yields in \snr\ well,
whereas the fewer ignition sparks result in Fe/IME ratios that
are too high.

\begin{figure}
  \centering 
  \includegraphics[angle=0, width=0.5\textwidth]{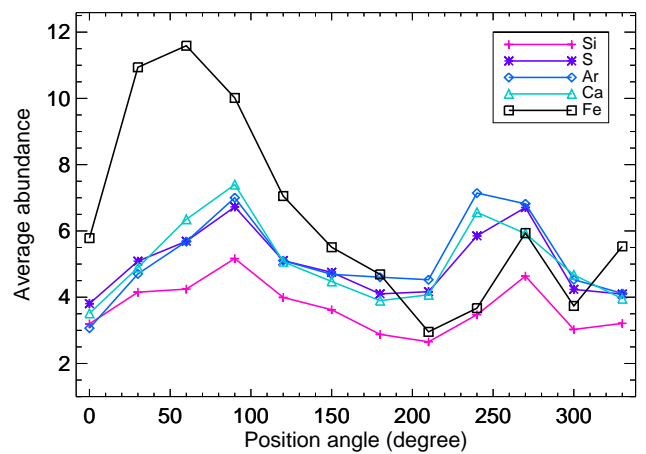}
\caption{
The average abundances as a function of the P.A. ($=0$ to the north;
counter-clockwise orientation). 
The abundance of each element is averaged inside the dashed circle 
denoted in Figure~\ref{F:density}.
}
\label{F:paabun}
\end{figure}

\subsubsection{Aspherical explosion} \label{S:aspherical}

The ejecta distribution in \snr\ is not spherical, as shown
in the abundance--position angle (P.A.) diagram (Figure~\ref{F:paabun}).
Here the explosion center is assumed at the approximate geometric center
$\alpha_{\rm J2000}=\RAdot{19}{11}{07}{6}$, $\delta_{\rm J2000}=
\decldot{09}{06}{10}{2}$, which is denoted by a
green cross in the bottom-right panel of Figure~\ref{F:pars2}.
In particular, the IGEs, synthesized in the densest part of the 
exploding WD, show strong lateral distribution.
The abundance of Fe is evidently enhanced in the eastern part of
\snr, and in the range 
P.A. $\sim 10$--$130^\circ$ (mean [Fe]$>6$).
The Si--Ca abundances are elevated in a similar P.A. range, but also
in a nearly opposite P.A. range ($\sim 210$--$300^\circ$).
The lateral distribution of Fe likely
reflects intrinsic asymmetries of the SN explosion.
The IMEs have a more axial symmetric morphology, which could be intrinsic to the explosion as well,
but as discussed in \S~\ref{S:bar-like} could also be caused by the structure of the circumstellar medium in which \snr\ is evolving.

A feasible explanation for an asymmetric Type-Ia SN explosion is off-center
ignition of a WD \citep{ropke07, maeda10b},
which has been used to interpret the spectral evolution diversity observed
in Type-Ia SNe \citep{maeda10b}.
\citet{maeda10a} modeled nucleosynthesis results of an extremely
off-center deflagration of a Chandrasekhar-mass WD followed
with a DDT (model ``O-DDT''), and compare them with those from a
spherically symmetric explosion (model ``C-DDT'').
As shown in the bottom-right panel of Figure~\ref{F:nucleo},
the off-center ignition model (29 sparks assumed) better describes 
the abundance ratios in \snr.
Furthermore, the off-center ignition model predicts an offset 
distribution of stable Fe-peak elements, which appears to explain 
the lateral Fe distribution.
The abundances and yields of two-dimensional simulations of an 
off-center SN explosion are identical to those of the 100 sparks ignition 
model as shown in the bottom-left panel of Figure~\ref{F:nucleo} \citep[see][for more comparison of
the two models]{seitenzahl13a},
and also similar to the nucleosynthesis results of a recent gravitational
detonation model with a single off-center bubble 
ignition \citep{seitenzahl16}.

Although the current study does not pin down 
a specific model among many Type-Ia models, we have found that \snr\ may provide
a unique and important example of an asymmetric Type-Ia explosion, which may offer more clues about
how WDs can explode asymmetrically.

\subsubsection{The high Mn abundance and its implications}

Most of the $^{55}$Mn is produced by the decay of $^{55}$Co via $^{55}$Fe, 
while the $^{55}$Co is mainly synthesized in 
the incomplete Si-burning and ``normal'' freeze-out from nuclear statistical equilibrium \citep[NSE;][]{seitenzahl13b}.
The production of the neutron-rich element 
$^{55}$Co (and hence $^{55}$Mn) requires the presence of 
neutron-rich elements \citep{badenes08, bravo13}. 
The abundance of these elements are influenced by several effects.
First of all, it may be caused by the presence of the initial 
abundance of the neutron-rich element $^{14}$N, in which case a 
high $^{55}$Mn yield may reflect the high initial metallicity 
of the progenitor  \citep{timmes03}. 
Using a one-dimensional Type-Ia explosion model 
(spherical, single ignition), \citet{badenes08} proposed a 
relation $M_{\rm Mn}/M_{\rm Cr}=5.3\times Z^{0.65}$
for estimating the progenitor $Z$ with the Mn and Cr supernova yields.
Secondly, in accreting C/O WDs  close to the 
Chandrasekhar limit, the rise in temperatures in the core will 
result in carbon fusion several centuries before the onset of 
the explosion. During this so-called carbon-simmering phase, 
weak interactions enhance the neutron fraction in the core 
\citep{piro08}, resulting in a higher $^{55}$Mn yield of the 
ensuing Type-Ia supernova. 
Finally, a high yield of $^{55}$Mn from  ``normal'' freeze-out from 
NSE requires that the 
nucleosynthesis occurs under conditions of high density
\citep[$\rho\gtrsim 2\E{8}\cm^{-3}$;][]{thielemann86}.
Such high density can be provided by WDs with 
masses larger than $1.2~\Msun$,  which is inconsistent 
with the sub-Chandrasekhar WD scenarios \citep{seitenzahl13b}.
But the high yield of Mn is consistent with the 
single-degenerate Type-Ia scenario, for which the 
mass  of the exploding WD should be close to the 
Chandrasekhar limit.
Even for Chandrasekhar-limit explosions, the Mn/Fe abundance ratio 
is rather sensitive to the explosion conditions.

For a few young SNRs, such as Tycho's and Kepler's SNRs, 
the high Mn/Cr ratio has been taken as evidence for a high initial
metallicity of the progenitor \citep{badenes08, park13},
since in these cases mostly the  outer layers of the supernovae were 
assumed to be shocked, and the $^{55}$Mn yield in the outer layer 
does depend on the  initial metallicity. 
Carbon simmering as the cause of the high Mn yield seems unlikely,
since it only affects the neutron excess of the core.
However, a high metallicity was suggested in the young SNR N103B based on metal ratios,  which seems highly unlikely given that this SNR is located in the low-metallicity  environment of the Large  Magellanic Cloud \citep{badenes16}.
In this case, but perhaps also in hindsight for the cases 
of Tycho's and Kepler's SNRs, the high yields may be affected 
by simmering, and radial mixing of material from the core and 
outer regions \citep[e.g.,][]{gamezo05}
may have affected the $^{55}$Mn abundance of 
outer ejecta \citep[but see][]{badenes05}. 
Finally, a large Mn/Cr ratio was also found in the Type-Ia 
SNR  candidate 3C~397, which could be contributed by the 
neutron-rich NSE region \citep{yamaguchi15}.
On the other hand, a dense, low-carbon WD with a solar- and 
subsolar-metallicity  progenitor can also produce the high Mn/Cr ratio in
3C~397 \citep{dave17}.

We report here\ a Mn mass of 
$M_{\rm Mn} =6.6\pm1.9\E{-3}\Msun$ for \snr, based on fitting the X-ray spectra
with the  $vrnei$ model. The mass ratio of Mn and Cr is $\sim 1.3$ (0.8--2.2).  
If we would attribute this to the effect of progenitor metallicity, a
super-solar metalicity of $Z=0.12_{-0.07}^{+0.14}=8^{+10}_{-4} Z_\odot$ 
is implied \citep[using
the solar ratio of][]{asplund09}.

\snr\ is in a later stage of its SNR evolution than young SNRs like Tycho's and Kepler's SNR,
so the Mn/Cr ratios reported here likely reflect 
the combination of both core and outer ejecta, that is,
from both ``normal'' freeze-out from NSE and incomplete 
silicon burning.
In this case, the ratio also
depends on the selected initial parameters such as 
central density and numbers of flaming bubbles \citep{seitenzahl13a} and the explosion mechanisms \citep[e.g.,][]{dave17}.

As shown in the bottom-left panel of Figure~\ref{F:nucleo},
the three-dimensional models with solar-metallicity progenitor and different
ignition sparks predict a range of $M_{\rm Mn}/M_{Cr}$ (0.6--2.2),
which overlaps with the mass ratio of 1.3 (0.8--2.2) reported here for \snr.
Therefore, from the point of view of existing models, and the SNR phase \snr\ in  
super-solar metallicity models may not be required
to explain our estimated high Mn/Cr mass ratio. 
Moreover, the high Mn/Cr ratio, and the high Mn/Fe ratio of  [Mn/Fe]/[Mn/Fe]$_\odot$ of 1.4--3.3
in \snr\ strongly suggest 
a WD explosion  almost the mass of Chandrasekhar
\citep[sub-solar Mn/Fe ratio for sub-Chandrasekhar cases;][]{seitenzahl13b}, 
which is best explained by the single-degenerate scenario.

\subsubsection{Properties for typing the SNR}

\begin{table}
\footnotesize
\caption{Properties for typing SNRs and \snr's results.}
\label{T:type}
\begin{tabular}{lcccc}
  \hline\hline
Properties & \snr\ & bipolar CC& normal CC & Type-Ia \\
\hline
PWN & no & + & -+ & + \\
NS & unclear & -+ & -+ & -+ \\
ejecta    & iron-group rich & - & - & +\\
asymmetry & yes  & + & + & -+\\
wind bubble  & $\lesssim 6$~pc & - & + & -+  \\
\hline
\end{tabular}
\tablefoot{The - and  + symbols under each explosion mechanism 
indicate negative and positive evidence, respectively, given
the properties of \snr\ listed in the second column.
The -+ symbols indicate that some known SNRs (not for all) with similar
properties match the scenario, or that the explosion mechanism cannot be constrained due to some unclear property.}
\end{table}

Table~\ref{T:type} summarizes the main properties for typing an SNR
and the results of \snr, in which their importance in reference is
reflected in the sequence.

\begin{enumerate}[i]
\item  {\sl PWN/NS}: There is no PWN detected inside \snr, or any clear evidence
of an associated NS, although the presence of
an NS cannot yet be excluded, as shown
by the presence of numerous point sources in and around \snr\ (Figure~\ref{F:ps}).
An association of one of them with an NS would strongly suggest
a CC origin.
\item  {\sl Ejecta}: The masses of the IGEs in the SNR are heavily overabundant compared to CC nucleosynthesis models. 
The observed abundance ratios instead suggest a Type-Ia origin.
\item {\sl Asymmetry}:
CC SNRs appear to be statistically more asymmetrical than 
Type-Ia SNRs \citep{lopez11}.
However, there is emerging evidence that a fraction of the Type-Ia SN 
explosion is intrinsically aspherical 
\citep[][for SN 1006]{maeda10b, uchida13}.
Moreover, 
the morphology of evolved SNR is subject to
the shaping by the inhomogeneous 
environment.
Given the nonuniform environment, Type-Ia SNe can also evolve to 
mixed-morphology SNRs.
For \snr, the centrally filled morphology is mainly due to density enhancement
in the SNR interior (see discussion below).
We note that also the candidate Type-Ia SNR 3C~397 
\citep[][the last reference suggests a CC origin]{chen99, yang13, yamaguchi14, yamaguchi15, safi-harb00} is highly aspherical, with many properties that are  similar to W49B.
\item {\sl Wind bubble}:
The presence and size of a wind bubble, and/or the stellar environment itself provides  clues about  the
progenitors of SNRs.
\snr\ is suggested to be located inside a wind-blown bubble,
whose relatively small size suggests that
the progenitor had a mass of $\sim13\Msun$, or smaller 
(if the SNR shock has already transgressed sufficiently into the shell)
if it is a CC SNR. 
However, Type-Ia SNRs can also be associated with wind bubbles 
if there is outflow driven by their single-degenerate progenitor systems, such as Tycho \citep{zhou16a}
and the Type-Ia candidate RCW~86 \citep{williams11,broersen14,badenes07}.
Moreover, observations of Type-Ia SNe have 
shown that a considerable population explodes ``promptly'' 
and  associated relatively young stellar populations, although the delay time of their 
explosions (40--420~Myr) is still larger
than that of CC SNe
\citep[$\lesssim 40$~Myr;][]{maoz12}. The presence of a small cavity surrounded by a dense ambient medium therefore either
suggests a CC SNR with a low-mass progenitor, or a Type-Ia
SNR, with a moderately sized wind-blown bubble.
\end{enumerate}

We favor the Type-Ia origin of \snr\ after a comparison of all direct 
and indirect properties for SNR typing, 
although some of its indirect properties can also be explained by
a normal CC explosion.
Both observations and models indicate that there is some diversity among 
Type-Ia SNe. In that regard,
\snr\ is an intriguing object revealing some properties different
from other Type-Ia SNRs. Further studies, especially on the details
of the explosion process and its environment, are needed to achieve
an integrated understanding of this remnant. 

\subsection{``Bar-like'' morphology due to a density enhancement}
\label{S:bar-like}

The bar-like X-ray morphology, which has an east-west orientation and is sharper in Fe-K emission,
has been attributed to either an asymmetric SN explosion origin \citep{miceli06, lopez13a}, 
 the density distribution of the ambient medium \citep{miceli06} shaped by the progenitor winds 
\citep{keohane07,miceli08,zhoux11} or to two opposite SN jets perpendicular to the bar-like structure \citep{bear17}.

Our density maps show that the cold component reveals a 
high-density distribution that has a similar east-west orientation, which is less prominently present in the
hot component. The strong association with the cold component,
which is likely shocked ambient medium, suggests that the bar-like structure is most likely the result of a structure in the ambient medium.

However, the X-ray morphology is clearly a combination of
the metal-rich ejecta distribution and the density distribution.
A possible explanation is that \snr\ evolves inside a more
or less barrel-shaped cavity, with lower densities in the North and South, which agrees well with the shape
of the infrared \citep{keohane07} and radio emission \citep{moffett94}.
The higher densities in the equatorial direction are reflected in the higher densities of the cold component along the bar. 
The high densities along the bar
probably also triggered the early formation of the reverse shock, and therefore also resulted
in the ejecta being shocked at higher densities in the central region of \snr. 

The intrinsic asymmetry of the ejecta is reflected not in the overall bar-like morphology of the X-ray line emission, but in the abundance maps (Figure~\ref{F:pars2}), which indeed reveal
not so much a bar-like morphology, but a fan-like region that
reveals higher Fe abundances in the Eastern part of the bar-like region.

A bipolar explosion jet seems unlikely, since the base of the
jet should be displaced several thousand years after the explosion \citep[see the numerical simulation in][]{gonzalez-casanova14}. Instead the bar-like region goes through the center
of the \snr, which is why this SNR is labeled a mixed-morphology SNR.

\subsection{Recombining gas and SNR age} \label{S:age}

As mentioned in Section~\ref{S:twocomp}, the recombining plasma occupies the
majority of the SNR regions of \snr\ ($\sim 3/4$ by area), not only 
in the southwest, but also 
in the bar-like structure in the SNR interior and in the northeast.
The recombination age in the  
northeast is larger than in
the southwest (see Figure~\ref{F:density}).
The gas that has reached CIE is distributed in the southeastern shell and 
spreads in patches throughout  the northeastern part of the remnant.

The distribution and origin of the recombining/over-ionized plasma in \snr\ has 
been studied in several papers, and rapid cooling is regarded as the cause
\citep[e.g.,][]{ozawa09}.
Thermal conduction was initially proposed as the cooling mechanism \citep{kawasaki05}, but 
adiabatic expansion is more favored in the later studies
\citep{miceli10, lopez13b}.
\citet{zhoux11} performed a hydrodynamic simulation of the \snr\ evolution in a
circumstellar dense cylinder as indicated by the infrared observations.
They found that the recombining plasma originates in both cooling processes:
(1) the mixing of hot plasma and cold gas evaporated from the dense circumstellar
medium; and (2) rapid adiabatic expansion.
The density enhancement along the bar-like structure (see Figure~\ref{F:pars2}) 
supports the existence of the dense circumstellar matter 
for process (1) to occur.  
The adiabatic expansion can explain the large-scale recombining plasma 
extended to the southwestern boundary.

The recombination age shown in Figure~\ref{F:density} is derived from the 
recombination timescale and the electron density in the hot phase ($t_{\rm r}=\tau_{\rm r}/ n_e$,
where the density is assumed to be a constant with time).
It describes the time elapse starting from the moment that plasma
reached ionization equilibrium, and then began cooling.
Our measurements indicate recombination ages between $\sim 2000$~yr and $\sim 6000$~yr, based on an over-ionization model. The true
age of the plasma may well differ as the ionization state
depends on the temperature and density history of the plasma. But one should also note that over-ionization can only occur
once ionization equilibrium has been reached, and the plasma
turns from under-ionized into over-ionized \citep{kawasaki05}.
The latter would imply that our recombination age measurements even underestimate the true plasma age. Although the connection between recombination age and SNR age is complex, the recombination ages that we find have
implications for the age of W49B.

In previous studies, the age of W49B was estimated
to be between $\sim 1000$~yr \citep{pye84,lopez13a} and $\sim 4000$~yr \citep{hwang00}. 
The recombination ages measured by us suggest that the SNR is older, 
as the SNR cannot be younger than the oldest plasma it contains, and the 
recombination age probably underestimates the true age of the plasma. 
Moreover, for the plasma to first reach ionization equilibrium, a timescale of 
$\gtrsim 10^{12} \cm^{-3} \s$ \citep{smith10} is needed. Even for densities around $n_\mathrm{H}\approx 10$~cm$^{-3}$, typical for the Western region, the associated timescale is $\gtrsim 2600$~yr. The recombination timescale of $\sim 2000$~yr for that region should be added to that. Therefore, the total plasma age must be of the order of 4000-5000~yr. In the eastern regions, the recombination
ages are as high as 6000~yr, or the plasma is in equilibrium (white pixels in Fig.~\ref{F:density}). Since the density is lower in that region ($n_\mathrm{H}\approx 4$~cm$^{-3}$), CIE implies plasma ages of $\sim 6600$~yr, close
to maximum recombination ages we find.
The recombination ages therefore suggest an age of at least 
5000--6000 yr.

One can also estimate the age based on the Sedov model combined
with an estimate of the forward shock velocity based on the plasma temperature. Here one has the choice between taking the hot
plasma component or the cool plasma component. However,
the hot plasma component seems to be very much metal enriched, 
and shows a lot of temperature variation. This component is likely
associated with plasma shocked by the reverse shock. Associating
the cooler component with the forward shock, we infer
a forward shock velocity of 
$v_{\rm s}=[16 kT_{\rm c}/(3\bar{\mu} m_{\rm H})]^{1/2} =476 (kT_{\rm c}/0.27~ {\rm keV})^{1/2} \km\ps$, 
where the mean atomic weight $\bar{\mu}=0.61$ is taken for fully ionized plasma and 
$m_{\rm H}$ is the hydrogen atomic mass.
For a uniform ambient medium,
the velocity corresponds to a Sedov age of 
$t_{\rm sedov}=2\Rs/(5v_{\rm s})\sim 5.3 (kT_{\rm c}/{\rm 0.27~ keV})^{-1/2}$~kyr, 
and an explosion energy $E_0=25/(4\xi)(1.4n_0 m_{\rm H})\Rs^3 v_{\rm s}^2\sim 1.3\E{51}n_{10} \erg$ \citep{ostriker88},
where we adopt an SNR radius $\Rs=6.4$~pc, and $\xi=2.026$ , with $n_{10}$ the ambient density of
hydrogen atom in units of 
$10~\cm^{-3}$.
The temperature of the cool component is therefore also consistent
with an age of 5000--6000~yr, and the canonical
supernova explosion energy of $\sim 10^{51}$~erg 
for the high
density environment in which W49B expands.

\section{Conclusion}

We have performed a spatially resolved X-ray study of SNR \snr\ using a state-of-the-art adaptive binning method
in order to uncover its explosion mechanism and the origin of the 
centrally filled morphology.
An asymmetric Type-Ia explosion is the most probable 
explanation for the abundances, yields, and metal distribution in \snr.
A density enhancement with an east-west orientation is the main reason for the bar-like X-ray morphology.
The detailed results are summarized as follows.

\begin{enumerate}
\item 
The X-ray emission in \snr\ is well characterized by two-temperature gas containing
a cool component with $\kTc\sim 0.27$~keV and a hot, ejecta-rich
component with $\kTh\sim 0.6$--2.2~keV. There is a large
gradient of $\kTh$ from the northeast to the southwest.
The detailed distribution of gas temperature and other physical parameters across the SNR are
shown (see Figure~\ref{F:pars2} and \ref{F:density}).

\item \snr\ is evolving in a dense environment. 
The mean densities are $24 \cm^{-3}$ and $5 \cm^{-3}$ for the cool and hot
X-ray components, respectively.
The total masses of the X-ray-emitting gas 
are $484^{+41}_{-34}\du^{2.5} \Msun$
in the cool phase and $52\pm8 \du^{2.5}\Msun$ in the hot phase.

\item We obtained the mass-weighted average abundances 
[Si]=$3.4^{+0.8}_{-0.7}$, [S]=$4.9_{-1.0}^{+1.1}$, 
[Ar]=$5.1_{-1.0}^{+1.2}$, [Ca]=$5.2_{-1.0}^{+1.2}$, and [Fe]=$5.7_{-1.1}^{+1.2}$.
The Cr, Mn, and Fe abundances according to a fit 
to the global spectra of \snr\ are 
[Cr]=$6.6\pm0.8$, [Mn]=$12.5\pm 2.7$,
and [Fe]=$3.2\pm 0.1$.

\item The element Fe  shows a strong lateral distribution.
The abundance of Fe is evidently enhanced in SNR east and in P.A. range $\sim 10$--$130^\circ$.
The Si--Ca abundances are also elevated in a similar P.A. range, but also 
in the nearly opposite P.A. range ($\sim 210$--$300^\circ$).
The lateral distribution of Fe  suggests intrinsic 
asymmetries of the SN explosion. The  nearly axially symmetric 
distribution of IMEs may also reflect that the explosion was not sphericially symmetric,  but in this case also the density distribution of the circumstellar medium may play a role (see point 8 below).

\item We have found 24 point-like sources in the vicinity of \snr\ with luminosities in the range $10^{32.1-36.2} \du^2 \erg\ps$.
Nine of them have the luminosities of a cooling NS at the age of \snr\ (at an assumed distance of 10 kpc).
Therefore, even if \snr\ is a CC SNR, there is still
the possibility that it contains a cooling NS,
rather than a black hole. 

\item 
None of the CC nucleosynthesis models 
(spherical explosion or bipolar explosion)
explain the abundance ratios in \snr. The iron-group yields predicted
by the CC models are insufficient to explain the observed masses in
the hotter phase:
$M_{\rm Cr}=4.9\pm 1.1 \E{-3}~\Msun$, 
$M_{\rm Mn}=6.6\pm 1.9\E{-3}~\Msun$, 
$M_{\rm Fe}=0.32^{+0.10}_{-0.09}~\Msun$. 
The energetic CC explosion scenario matches even more poorly than 
the normal CC scenario given the small molecular cavity surrounding the SNR and nine
point-like sources in the vicinity of \snr\ (probably only a projection effect).

\item A DDT Type-Ia model with multi-spot ignition of a 
Chandrasakhar-mass WD well describes the observed abundance ratios.
This model based on solar-metallicity can also explain the 
high Mn-to-Cr ratio ($M_{\rm Mn}/M_{\rm Cr}= 0.8-2.2$) found in W49B.
A feasible explanation of the asymmetric Type-Ia SN explosion 
is off-center ignition of a WD.

\item
The centrally filled/bar-like morphology of \snr\ is mainly due to
density enhancement projected to the SNR center ($n_c >30\cm^{-3}$), given
the good spatial correlation between the gas density
and the X-ray brightness.  This suggests that \snr\ evolves in a barrel-shaped cavity, which also lead the ejecta to be shocked
at higher densities and projected to the center.
The overall morphology, triggered by the ambient medium structure, combined with our conclusion that \snr\ is a Type-Ia SNR, suggests that Type-Ia supernovae can also result in mixed-morphology SNRs.

\item  The recombination age of the plasma suggests 
an SNR age of $\sim 5$--6 kyr,
similar to the estimated Sedov age of 5.3 kyr.

\end{enumerate}

\begin{acknowledgements}
P.Z. acknowledges the support from the NWO Veni Fellowship, grant no. 639.041.647
and NSFC grants 11503008, 11590781, and 11233001.
\end{acknowledgements}

\bibliographystyle{aa} 
\bibliography{w49b}

\appendix

\section{Point-like sources in the vicinity of \snr}

\begin{table*}
\footnotesize
\caption{Information of the point-like sources in the vicintiy of W49B}
\label{T:ps}
\begin{tabular}{lccccccccccc}
  \hline\hline
No.
& \multicolumn{2}{c}{Coordinates} & \multicolumn{2}{c}{counts (hard)} & \multicolumn{2}{c}{counts (soft)} & significance
& photon flux $f$ & Hardness ratio  & $kT_{\rm BB}$ & $\log_{10}(L_{\rm BB})$\\
\cmidrule(lr){2-3} \cmidrule(lr){4-5} \cmidrule(lr){6-7}
& R.A. & Dec. & total & net  & total & net  & ($\sigma$) & (photon s$^{-1}$ cm$^{-2}$) 
& $\frac{(f_{\rm H}-f_{\rm S})}{(f_{\rm H}+f_{\rm S})}$ & (keV) & ($\erg \ps)$ \\
\hline
           1
  & 19:11:05.7
  & 09:02:33.6
  & 75
  & 62.5
  &191
  &163.5
  & 29.3
  &  1.8E-06$\pm$  5.8E-08
  &-0.24$\pm$ 0.02
  & 0.25
  &34.2
\\
           2
  & 19:11:14.6
  & 09:03:28.5
  & 17
  &  0.9
  &126
  & 73.8
  & 11.0
  &  4.5E-07$\pm$  3.9E-08
  &-0.96$\pm$ 0.08
  & 0.12
  &36.2
\\
           3
  & 19:11:21.2
  & 09:05:02.7
  & 26
  & 12.9
  &226
  &168.8
  & 17.8
  &  1.2E-06$\pm$  4.6E-08
  &-0.79$\pm$ 0.03
  & 0.17
  &35.2
\\
           4
  & 19:11:06.1
  & 09:05:33.4
  &166
  & 75.4
  & 90
  &  6.9
  &  5.4
  &  7.4E-07$\pm$  5.1E-08
  & 0.89$\pm$ 0.05
  & 1.18
  &32.4
\\
           5
  & 19:10:57.1
  & 09:05:39.5
  & 35
  & 23.7
  & 22
  & -0.4
  &  4.8
  &  2.2E-07$\pm$  3.5E-08
  & 1.00$\pm$ 0.11
  & $>5$
  & -
\\
           6
  & 19:10:50.5
  & 09:07:24.7
  & 29
  & 22.2
  & 24
  & 13.3
  &  9.1
  &  6.8E-07$\pm$  8.3E-08
  & 0.44$\pm$ 0.08
  & 0.42
  &32.8
\\
           7
  & 19:11:04.0
  & 09:07:37.6
  &155
  & 20.6
  &338
  &116.4
  &  9.1
  &  1.0E-06$\pm$  6.7E-08
  &-0.54$\pm$ 0.06
  & 0.21
  &34.4
\\
           8
  & 19:10:57.4
  & 09:04:02.1
  &  1
  & -1.0
  & 35
  & 24.5
  &  5.8
  &  2.2E-07$\pm$  3.6E-08
  &-1.00$\pm$ 0.14
  & $<0.01$
  & -
\\
           9
  & 19:11:04.2
  & 09:04:04.0
  & 21
  &  1.0
  & 57
  & 22.9
  &  6.2
  &  1.4E-07$\pm$  3.6E-08
  &-0.87$\pm$ 0.27
  & 0.15
  &34.7
\\
          10
  & 19:11:10.6
  & 09:06:37.3
  &491
  & 69.3
  &244
  & 51.2
  &  5.2
  &  9.9E-07$\pm$  7.0E-08
  & 0.38$\pm$ 0.04
  & 0.40
  &33.1
\\
          11
  & 19:11:21.8
  & 09:06:43.0
  &  9
  &  5.2
  & 12
  &  4.5
  &  3.7
  &  7.8E-08$\pm$  2.8E-08
  & 0.30$\pm$ 0.24
  & 0.37
  &32.1
\\
          12
  & 19:11:07.5
  & 09:06:54.9
  &113
  & 21.5
  & 87
  & -4.4
  &  3.3
  &  2.1E-07$\pm$  4.9E-08
  & 1.00$\pm$ 0.17
  & $>5$
  &  -
\\
          13
  & 19:11:08.1
  & 09:07:27.0
  & 90
  & 36.1
  & 63
  & 10.4
  &  3.9
  &  4.3E-07$\pm$  4.9E-08
  & 0.71$\pm$ 0.07
  & 0.60
  &32.3
\\
          14
  & 19:11:09.9
  & 09:09:04.4
  & 31
  & 10.1
  & 77
  & 53.3
  &  9.3
  &  4.1E-07$\pm$  4.1E-08
  &-0.52$\pm$ 0.09
  & 0.21
  &34.0
\\
          15
  & 19:11:05.3
  & 09:02:53.2
  &  1
  &  0.1
  &  7
  &  5.3
  &  2.3
  &  5.6E-08$\pm$  3.3E-08
  &-0.95$\pm$ 0.58
  & 0.13
  &34.9
\\
          16
  & 19:11:14.1
  & 09:03:08.9
  & 31
  & 21.0
  & 22
  &  1.2
  &  5.5
  &  2.1E-07$\pm$  3.5E-08
  & 0.93$\pm$ 0.12
  & 1.83
  &32.1
\\
          17
  & 19:11:18.9
  & 09:04:03.8
  & 10
  & -2.0
  & 54
  & 23.6
  &  5.1
  &  1.5E-07$\pm$  3.4E-08
  &-1.00$\pm$ 0.23
  & $<0.01$
  & -
\\
          18
  & 19:11:12.5
  & 09:05:32.4
  &298
  & 27.5
  &400
  & 78.6
  &  5.0
  &  7.1E-07$\pm$  6.5E-08
  &-0.27$\pm$ 0.07
  & 0.25
  &33.8
\\
          19
  & 19:10:59.2
  & 09:05:48.6
  &180
  & 51.1
  &136
  & 24.7
  &  5.2
  &  6.2E-07$\pm$  5.4E-08
  & 0.54$\pm$ 0.05
  & 0.47
  &32.7
\\
          20
  & 19:11:13.0
  & 09:06:09.3
  &307
  & 15.3
  &342
  & 60.7
  &  4.7
  &  5.0E-07$\pm$  6.5E-08
  &-0.41$\pm$ 0.12
  & 0.23
  &33.9
\\
          21
  & 19:11:10.1
  & 09:06:56.0
  &197
  & 12.2
  &196
  & 53.0
  &  4.5
  &  4.9E-07$\pm$  6.6E-08
  &-0.45$\pm$ 0.13
  & 0.22
  &34.0
\\
          22
  & 19:11:05.9
  & 09:07:30.9
  &245
  & 90.0
  &221
  & 52.7
  &  8.5
  &  1.4E-06$\pm$  7.0E-08
  & 0.48$\pm$ 0.03
  & 0.44
  &33.1
\\
          23
  & 19:11:07.9
  & 09:10:19.7
  &  1
  & -0.7
  & 19
  & 13.0
  &  3.6
  &  9.7E-08$\pm$  2.8E-08
  &-1.00$\pm$ 0.25
  & $<0.01$
  & -
\\
          24
  & 19:11:15.7
  & 09:10:26.3
  &  4
  &  1.9
  & 11
  &  7.9
  &  3.5
  &  8.6E-08$\pm$  3.3E-08
  &-0.43$\pm$ 0.32
  & 0.22
  &33.2
  \\
  \hline
\end{tabular}
\end{table*}

We detected 24 point-like sources in the 0.7--5~keV energy band 
in the vicinity of \snr\ using the Mexican-Hat wavelet source detection 
tool $wavdetect$ in CIAO (see Figure~\ref{F:ps}). 
We combined the three Chandra observations to detect faint sources,
where the exposure-weighted point spread function map was use
to run $wavdetect$ on the merged dataset\footnote{http://cxc.harvard.edu/ciao/threads/wavdetect\_merged}.
The energy band is selected to highlight the emission of the 
point-like sources over the SNR background. 
The compact object is assumed to have an effective temperature of 
0.07--0.5 keV for heavy element atmosphere or 0.11--0.5 keV for 
Hydrogen atmosphere, with a foreground absorption column density 
5.5--8.8$\E{22}~\cm^{-2}$.
Two criteria are considered for the NS temperature range: 
1) The observed central compact objects show blackbody temperatures in 
the range 0.2--0.5~keV \citep{pavlov04} and
2) the minimal paradigm \citep{page04} predicts a temperature of 0.11 keV for 
the light elements envelop and 0.07 keV for the heavy envelop of
an NS with an age of 6000 yr.

To minimize the contamination from the SNR structures, we extracted 
the most compact sources using the detection spatial scales of 1, 
$\sqrt{2}$, and 2 pixels.
Adding a larger spatial scale of $2\sqrt{2}$ would result in 
a detection of 76 sources, while most of these sources are likely 
clumpy plasma inside the SNR. 
We examined the detected sources in the hard (2.5--5 keV) and soft (0.7--2.5 keV)
in order to study their spectral properties.
Table~\ref{T:ps} summarizes the coordinates, total (background-included) and net 
(background-subtracted) counts in the hard and soft bands, photon fluxes $f$, 
and hardness ratios $HR$ of the sources, where $f$ is the net count rate divided 
by the effective area of the camera.
The hardness ratio defined as $HR=(f_{\rm H}-f_{\rm S})/(f_{\rm H}+f_{\rm S})$ 
has a value between -1 and 1, where
the $f_{\rm H}$ and $f_{\rm S}$ are the photon fluxes in the hard and soft bands, 
respectively.

The blackbody temperatures $kT_{\rm BB}$ and luminosities $L_{\rm BB}$ of the 
point-like 
sources are estimated using hardness ratio $HR$ and photon flux $f$
(see Figure~\ref{F:hr_8e22}).
The foreground absorption and distance are assumed to be $\NH = 8\E{22} \cm^{-2}$ 
and 10~kpc, respectively, the same as those of \snr.
The best-fit $kT_{\rm BB}$ and $L_{\rm BB}$ are also listed in Table~\ref{T:ps}.

\end{document}